\documentclass[pre, aps,twocolumn,floatfix,superscriptaddress]{revtex4-1}
\usepackage{amsmath,amssymb,graphicx,cases}
\usepackage{epsfig}
\usepackage{textcomp}
\usepackage{epstopdf}
\usepackage{float}
\usepackage{braket}
\usepackage[normalem]{ulem}

\usepackage[normalem]{ulem}
\newcommand{\stkout}[1]{\ifmmode\text{\sout{\ensuremath{#1}}}\else\sout{#1}\fi}

\usepackage{fixmath}
\newcommand{\vect}[1]{\mathbold {#1}} % for vectors
\usepackage{color}
\definecolor{Blue}{rgb}{0.00, 0.00, 1.00}
\definecolor{Red}{rgb}{1.00, 0.00, 0.00}
\definecolor{Green}{rgb}{0.00, 0.60, 0.00}

\newcommand{\nn}{\nonumber}
\newcommand{\be}{\begin{equation}}
\newcommand{\ee}{\end{equation}}
\newcommand{\bea}{\begin{eqnarray}}
\newcommand{\eea}{\end{eqnarray}}

\usepackage[colorlinks=true, urlcolor=blue, anchorcolor=blue, citecolor=blue,filecolor=blue,linkcolor=blue,menucolor=blue]{hyperref}

\begin{document}
\title{Nonequilibirum steady state for harmonically-confined active particles}

\author{Naftali R. Smith}
\email{naftalismith@gmail.com}

\affiliation{Department of Solar Energy and Environmental Physics,
  Blaustein Institutes for Desert Research, Ben-Gurion University of
  the Negev, Sede Boqer Campus, 8499000, Israel}

\author{Oded Farago}

\affiliation{Department of Biomedical Engineering, Ben-Gurion
  University of the Negev, Marcus Family Campus Be'er Sheva, 8410501,
  Israel}


\begin{abstract}

We study the full nonequilibirum steady state distribution
$P_{\text{st}}\left(X\right)$ of the position $X$ of a damped particle
confined in a harmonic trapping potential and experiencing active
noise, whose correlation time $\tau_c$ is assumed to be very
short. Typical fluctuations of $X$ are governed by a Boltzmann
distribution with an effective temperature that is found by
approximating the noise as white Gaussian thermal  noise. However, large deviations of $X$ are described by a
non-Boltzmann steady-state distribution. We find that, in the limit
$\tau_c \to 0$, they display the scaling behavior
$P_{\text{st}}\left(X\right)\sim e^{-s\left(X\right)/\tau_{c}}$, where
$s\left(X\right)$ is the large-deviation function.  We obtain an
expression for $s\left(X\right)$ for a general active noise, and
calculate it exactly for the particular case of telegraphic
(dichotomous) noise.

\end{abstract}

\maketitle

\section{Introduction}

\subsection{Background}

One of the fundamental problems in the study of non-equilibrium
statistical mechanics and probability theory is fluctuations in
stochastic systems. Rare events, or large deviations, are particularly
interesting \cite{Varadhan,O1989,DZ,Hollander,Bray, Majumdar2007,
  hugo2009,Derrida11, RednerMeerson, Vivo2015, MeersonAssaf2017,
  MS2017,Touchette2018}. Despite their small likelihood, they are of great interest because of their relevance to catastrophes like
%can have devastating consequences. Examples include
earthquakes, heatwaves, population extinction, and stock-market
crashes. It is therefore important to develop theoretical frameworks
that enable one to estimate their occurrence probability.
%for them to occur.
Here, we study statistics of rare events within the context of
%In recent years there is a growing interest in
\emph{active} stochastic systems, which physically correspond to
molecules and objects that consume energy from their environment and
convert it into directed motion \cite{Mizuno07, Wilhelm08, Bruot11,
  Ahmed15, Breoni22}. Such systems are driven out of thermal
equilibrium, and their dynamics may exhibit a multitude of
counter-intuitive and rich phenomena \cite{Marchetti13, ALP14, FodorEtAl15,
  Kardar15, Dhar_2019, Fodor16, Needleman17, MalakarEtAl18, led19, WZKLT19, WKL20, 
  led20, Basu20, SBS20, MB20, LMS21, ARYL21, WHB22, FJC22, SBS22, ARYL22, ACJ22}.  In
particular, they can relax to nonequilibrium steady states (NESSs),
which in general cannot be described by a Boltzmann distribution.
%(in fact, one cannot even define the temperature or pressure of active systems in general).

The first-passage time (FPT) problem (or Kramers' escape problem) is a
classical problem in statistical mechanics \cite{RednerFPTBook, BMS13}.  In
the simplest setting, one considers a particle that is trapped in a
potential well and interacts with a noisy environment. The position
$x$ of the particle at time $t$ is described by a Langevin equation,
which for a system in one-dimension reads
\be
\label{Langevin}
m\ddot{x}+\Gamma\dot{x}+V'\left(x\right)=\xi\left(t\right)\, , 
\ee
where $m$ is the particle's mass, $\Gamma > 0$ is the damping
coefficient, $V(x)$ is the external potential, and $\xi(t)$ is the
noise.  The FPT problem is that of determining the distribution of the
first time $t_{\text{esc}}\left(X\right)$ at which $x(t)$ reaches some
target $X$, given an initial condition. A related interesting quantity
to study is the steady-state distribution (SSD), $P_{\text{st}}(X)$,
of the position of the particle.

If $\xi(t)$ describes thermal noise, then it can be mathematically
described by a white Gaussian noise with the following statistical
properties: $\left\langle \xi\left(t\right)\right\rangle =0$, and,
$\left\langle \xi\left(t\right)\xi\left(t'\right)\right\rangle
=2k_{B}T\Gamma\delta\left(t-t'\right)$, where angular brackets denote
the expectation value, and $k_B$ and $T$ denote Boltzmann's constant
and the temperature, respectively.  For thermal noise, the SSD is
given by the Boltzmann distribution, 
\be
\label{BoltzmannDist}
P_{\text{st}}\left(X\right)\propto e^{-V\left(X\right)/k_{B}T} .  \ee
In the limit of low temperature and/or high barrier
  \cite{footnote:SM20}, the mean first passage time (MFPT) is
approximately given by Kramers’ formula (the Arrhenius law)
\cite{Kramers40, Hanggi90, Melnikov91} \be
\label{Arrhenius}
\left\langle t_{\text{esc}}\left(x=X\right)\right\rangle \sim e^{\Delta
  E/k_{B}T},
\ee
where $\Delta E$ is the difference between the potential energy at
positions $X$ and at the minimum of the potential well, i.e., $\Delta
E=V\left(X\right)-V_{\min}$.

A simple model of \emph{active} particle dynamics is given by
Eq.~\eqref{Langevin} with $\xi(t)$ that is not a thermal noise.  In
Refs.~\cite{BenIsaac15, Wexler20, Woillez20, GP21}, a model was
introduced and studied that is mathematically described by
Eq.~\eqref{Langevin} with a harmonic trapping potential,
$V\left(x\right)=kx^{2}/2$, and $\xi(t)$ that is the sum of a thermal
noise and an active (non-thermal) noise with correlation time
$\mathcal{T}_c$.
%
%This model describes the dynamics which, at a microscopic level, are driven by molecular motors. 
The model was originally inspired by experiments in ``active gels'',
which consist of a network of actin filaments and myosin-II molecular
motors in which the athermal nature of the random motion was probed by
measuring the fluctuations of tracked tracer particles, and showing
the fluctuation-dissipation theorem breaks down \cite{BHG06, Toyota11,
  Stuhrmann12, Mizuno07}. Position and velocity distributions and FPTs
were studied \cite{BenIsaac15, Wexler20, Woillez20}, as well as the
entropy production \cite{GP21}, but analytic progress could only be
made in certain limiting cases, such as the limits of very large or
very small $\mathcal{T}_c$.  In the latter case, the MFPT was found to
be given by Eq.~\eqref{Arrhenius}, but with a modified, effective
temperature $T_{\text{eff}}$, i.e., \cite{Wexler20}
\be
\label{Teff}
\left\langle
t_{\text{esc}}\left(x=X\right)\right\rangle \sim
e^{kX^{2}/2k_{B}T_{\text{eff}}}.
\ee

Here, we consider the case where $\xi(t)$ is a pure active noise
(i.e., no thermal component \cite{footnote:thermalNoise}) in the limit
$\mathcal{T}_c \to 0$ at a \emph{fixed} noise amplitude. This case
%It is important to
  has to be distinguished from the case where simultaneously one also
takes the noise amplitude to infinity with a proper
scaling. In the latter, the correlation function of the
noise becomes a delta function $\left\langle
\xi\left(t\right)\xi\left(t'\right)\right\rangle=2k_{B}T_\text{eff}\Gamma\delta\left(t-t'\right)$,
%for
with some effective temperature $T_\text{eff}$, and
Eq.~\eqref{Teff} becomes exact \cite{colored}.
In the former, as we will show below, Eq.~\eqref{Teff}
  (with $T_\text{eff}\propto\mathcal{T}_c$)  faithfully describes the
behavior of the system {\em only}\/ for small $X$ which
is sufficiently near the center of the trap.
%and in general, it
This result, however, breaks down at large $X$.  One way
to see this is to consider as an example noise terms that are bounded,
e.g., a telegraphic active noise.
%(and assume that there is no additional thermal noise term). In this case,
For this kind of noise $x(t)$ is bounded too, leading to a divergence of the
MFPT at some finite $X$, in contradiction with Eq.~\eqref{Arrhenius}
(regardless of the definition of $T_{\text{eff}}$).
%This is the case on which we will focus in the present work. Thus we find
This implies that, despite the timescale separation
present in the system, detailed balance is violated. A similar
situation was observed recently in a different context \cite{HG21}.

%\subsection*{\titlesize Project description}

\subsection{Precise definition of the problem, rescaling, and summary of main results}

Let us consider the Langevin equation \eqref{Langevin} in the
particular case of a harmonic trapping potential
$V\left(x\right)=kx^{2}/2$,
\be
\label{LangevinDimensional}
m\ddot{x}\left(t\right)+\Gamma\dot{x}\left(t\right)+kx\left(t\right)=\Sigma\left(t\right),
\ee
where $\Sigma\left(t\right)$ is telegraphic (dichotomous) noise,
which switches between the values $\pm\Sigma_{0}$ at a constant rate
$1/\mathcal{T}_{c}$, i.e., the time between every two consecutive
switches is exponentially distributed with mean $\mathcal{T}_{c}$.
Let us first rescale space and time, $kx/\Sigma_{0}\to x$,
$\sqrt{k/m}\,t\to t$, so Eq.~\eqref{LangevinDimensional} becomes
\be
\label{LangevinDimensionless}
\ddot{x}+\gamma\dot{x}+x=\sigma\left(t\right) \, ,
\ee
where $\gamma=\Gamma/\sqrt{mk}$ and $\sigma\left(t\right)$ takes the
values $\pm1$, and the rescaled mean time for switching between these
two values is $\tau_{c}=\sqrt{k/m}\,\mathcal{T}_{c}$.  At long times,
the system described by Eq.~\eqref{LangevinDimensionless} will
approach a steady state.  Due to the active nature of the noise term,
which violates detailed balance, this will be a nonequilibrium steady
state.  Our goal is to calculate the full SSD
$P_{\text{st}}\left(X\right)$ of the particle's position, including
the tails of this distribution.
%, and a related quantity, the distribution of the FPT $t_{\text{esc}}\left(X\right)$ (for the FPT, we assume an initial condition that is located near the center of the trap, $x=\dot{x}=0$).
Due to the mirror symmetry of the problem,
$P_{\text{st}}\left(X\right) = P_{\text{st}}\left(-X\right)$ and the
distributions of $t_{\text{esc}}\left(X\right)$ and
$t_{\text{esc}}\left(-X\right)$ are identical, so we will only
consider $X>0$ in what follows.

This problem, as formulated above, presents a significant theoretical
challenge, and in order to make analytic progress, certain simplifying
assumptions must be made. In the strongly-overdamped limit $\gamma \to
\infty$, the SSD is known exactly for arbitrary trapping potentials
$V(x)$ \cite{q-optics1,q-optics2,q-optics3,q-optics4,Kardar15,
  Dhar_2019}.
%This result was recently extended to a noise term with three possible states for the harmonic oscillator \cite{Basu20}.
In this work, we consider Eq.~\eqref{LangevinDimensionless} with
arbitrary damping coefficient $\gamma$, and focus on the limit of
rapid switching rate of the noise term, $\tau_c \to 0$. In this limit,
the rapid switching of the noise causes it to typically average out to
zero over timescales that are relevant for the harmonic oscillator
(which are of order unity in our choice of units). Hence, it becomes
very unlikely to reach positions $X$ that are located away from the
center of the trap, so that the SSD is strongly localized around
$X=0$, and fluctuations become very rare.  In order to study the full
SSD (including its tails), it is thus natural to employ tools from
large-deviation theory.  In particular, we will combine the
optimal-fluctuation method (OFM) (or weak-noise theory)
\cite{Onsager,MSR, Freidlin,Dykman,Graham, Falkovich01, GF,EK04,
  Ikeda2015,Grafke15, bertini2015, Holcman, SmithMeerson2019,
  MeersonSmith2019, AiryDistribution20} with the Donsker-Varadhan (DV)
large-deviation formalism \cite{Bray, Majumdar2007,hugo2009,
  TouchetteMinimalModel, Touchette2018, Escamilla19,
  AiryDistribution20} in a novel way.

Let us summarize the principal results of this work.  We find that, in
the limit $\tau_c \to 0$, the SSD is given by
\be
  \label{PstScaling}
P_{\text{st}}\left(X\right)\sim e^{-s\left(X\right)/\tau_{c}}
  \ee
where the large-deviation function (LDF) $s(X)$ does not depend on
$\tau_c$.  Thus, reaching any $X$ that is sufficiently far from the
center of the trap is a large deviation.  It then follows (see Appendix \ref{app:MFPTandSSD}) that the FPT
to reach position $X$ follows an exponential distribution whose mean
is given (in the limit $\tau_c \to 0$) by 
%$\left\langle t_{\text{esc}}\left(X\right)\right\rangle
%\sim1/P_{\text{st}}\left(X\right)$, i.e., between the MFPT and the
%SSD is a very general result, holding for a very broad class of
%stochastic processes in the large-deviations regime. It leads to the
%large-deviation principle \eqref{SofXdef} for the MFPT, in terms of
%the same LDF $s(X)$.  and the FPT follows an exponential distribution
%whose mean is described by the same LDF, scaling as
\be
\label{SofXdef}
\left\langle t_{\text{esc}}\left(X\right)\right\rangle
\sim1/P_{\text{st}}\left(X\right)\sim e^{s\left(X\right)/\tau_{c}} \, .  
\ee
To be precise, we find a large-deviation principle (LDP)
\be
\label{LDPeq}
-\lim_{\tau_{c}\to0}\tau_{c}\ln
P_{\text{st}}\left(X\right)=\lim_{\tau_{c}\to0}\tau_{c}\ln\left\langle
t_{\text{esc}}\left(X\right)\right\rangle =s\left(X\right)\,.  \ee We
calculate $s(X)$ exactly, by finding the optimal (most likely)
trajectory of the particle, coarse-grained over timescales much longer
than $\tau_c$, that starts at the center of the trap at time $t =
-\infty$ and arrives at $x=X$ at time $t=0$.  The LDF $s(X)$ displays
a parabolic behavior around the center of the trap, at $X \ll 1$,
matching the description in terms of a Boltzmann distribution with an
effective temperature, as found in \cite{Wexler20}.  We find
intriguing qualitative differences between the system's behavior
between the overdamped ($\gamma > 2$) and underdamped ($\gamma < 2$)
cases. % ($\gamma > 2$ and $\gamma < 2$).
In particular, we
find that the SSD has a finite support $\left[-X_{0},X_{0}\right]$ (so
the motion of the particle is bounded and it cannot leave this
interval), where $X_0$ is a function of $\gamma$ that exhibits a
transition
%(i.e., a singularity)
at the critical value $\gamma = 2$.
  Finally, we verify our theoretical predictions through extensive
  computer simulations, and discuss extensions to general trapping
  potentials and other types of active noise.

\section{Theoretical framework}
\label{sec:theory}

Our theoretical framework consists of two main steps: (i) Providing an
effective coarse-grained description of the dynamics that is valid
over timescales much larger than $\tau_c$, which gives the probability
of (coarse-grained) trajectories of the particle, (ii) Calculating the
optimal history of the system that leads to a given $X$, i.e., the
most likely (coarse-grained) trajectory $x(t)$ that satisfies the
conditions $x(t \to -\infty) = 0$ and $x(t = 0) = X$.  The first step
relies on the well-established DV formalism, which we recall in
Appendix \ref{sec:DV}, while the second step involves the solution of
a minimization problem from the calculus of variations, and
essentially provides an extension of the OFM to active noise.

\subsection{Coarse graining}

In order to coarse grain the dynamics, let us consider the average of
the noise term over a timescale $t$ that is much longer than
$\tau_c$. We define
\be
\bar{\sigma}=\frac{1}{t}\int_{0}^{t}\sigma\left(\tau\right)d\tau \ee
Since the process $\sigma(t)$ is ergodic, its long-time averages
converge to their corresponding ensemble-average values. Therefore, in
the limit $ t / \tau_c \to \infty$, $\bar{\sigma}\left(t\right)$ will
approach its ensemble-average value which, for the telegraphic noise,
is zero.  Indeed, in the limit $\tau_c \to 0$ the noise effectively
becomes very weak, making fluctuations of $\bar{\sigma}$ very
unlikely.  Nevertheless, one can quantify the probability for such
fluctuations to occur. Fluctuations of integrals of stochastic
processes (known as ``dynamical'' or additive observables) over long
times have in fact been extensively studied, and there exists a
well-established theoretical framework for studying them. This theory,
sometimes referred to as Donsker-Varadhan (DV) theory, is based on a
path-integral representation of the process and uses the Feynman-Kac
formula \cite{Bray, Majumdar2007,
  DonskerVaradhan,Ellis,hugo2009,Touchette2018}.  The long-time
behavior predicted by DV theory is that of an LDP for the probability
density function (PDF) of the dynamical observable, describing an
exponential decay in time. In our particular case, it predicts that
the PDF $P\left(\bar{\sigma};t\right)$ of $\bar{\sigma}$ behaves as
%This behavior holds under quite general conditions, that are valid in a broad class of stochastic processes.
%Under quite general conditions, capturing a broad class of stochastic processes, the theory predicts that at long times the probability density function (PDF) $P\left(\bar{\sigma};t\right)$ of $\bar{\sigma}$ obeys a large-deviation principle (LDP) at long times:}
%\be
%\label{eq:DVscaling}
%P\left(\bar{\sigma};t\right)\sim e^{-t I\left(\bar{\sigma}\right)},\quad t\to\infty \, ,
%\ee
\be
\label{sigmabarLDP}
P\left(\bar{\sigma};t\right)\sim e^{-t\Phi\left(\bar{\sigma}\right)/\tau_{c}},
\ee
where the rate function $\Phi\left(\bar{\sigma}\right)$ is independent
of $t$ and $\tau_c$.  Eq.~\eqref{sigmabarLDP} holds at fixed $\tau_c$,
in the limit $t\to\infty$.

Importantly, by simply rescaling time, one finds that
Eq.~\eqref{sigmabarLDP} holds at \emph{fixed} $t$, in the limit
$\tau_c \to 0$.  The rate function $\Phi\left(z\right)$ that
corresponds to telegraphic noise has been found in several contexts,
such as run-and-tumble particles \cite{Dean21} and the Ehrenfest Urn
model \cite{MeersonZilber18}, and it is given by
\be
\label{Phidef}
\Phi\left(z\right)=1-\sqrt{1-z^{2}} \, , \ee
with $z$ confined to the interval $z\in\left[-1,1\right]$, as
obviously must be the case since $\sigma(\tau)$ is itself bounded so
that it cannot exceed this interval.  A calculation of
$\Phi\left(z\right)$, using DV theory, is given in Appendix
\ref{sec:DV} for completeness.

The coarse-graining procedure that we employ now is fairly standard
but we will, nevertheless, explain it now in detail. We consider a
sequence of time averages of the noise,
$\bar{\sigma}_{1},\bar{\sigma}_{2},\dots,\bar{\sigma}_{N}$, defined as
\be \bar{\sigma}_{i}=\frac{1}{\Delta
  t_{i}}\int_{t_{i}}^{t_{i+1}}\sigma\left(\tau\right)d\tau \, , \ee
for some sequence of $N$ time intervals, defined by
$t_{\text{initial}} \equiv t_{1}<t_{2}<\dots<t_{N+1} \equiv
t_{\text{final}}$, where $\Delta t_{i}=t_{i+1}-t_{i}$.  In the limit
$\tau_c \to 0$, the $\bar{\sigma}_{i}$'s become statistically
independent, so their joint PDF
$P\left(\bar{\sigma}_{1},\dots,\bar{\sigma}_{N}\right)$ is given by a
product of their individual PDFs, each of which is given by
Eq.~\eqref{sigmabarLDP}, i.e., \be
\label{JPDFsigmabar}
P\left(\bar{\sigma}_{1},\dots,\bar{\sigma}_{N}\right)\sim\exp\left[-\frac{1}{\tau_{c}}\sum_{i=1}^{N}\Delta
  t_{i}\Phi\left(\bar{\sigma}_{i}\right)\right] \, .  \ee The final
step in the coarse-graining procedure is to take the continuum limit
in \eqref{JPDFsigmabar} by treating the index $i$ of $\bar{\sigma}_i$
as continuous. This requires a separation of timescales, between the
short timescale $\tau_c$ and some longer characteristic timescale that
depends on the physical system under study.  We thus consider a
coarse-grained noise $\bar{\sigma}\left(t\right)$ whose distribution
is given by
%\bea
%\label{sigmaBarP}
%P\left[\bar{\sigma}\left(t\right)\right] &\sim& e^{-s\left[\bar{\sigma}\left(t\right)\right]/\tau_{c}}, \\	
%\label{sigmaBarS}
%s\left[\bar{\sigma}\left(t\right)\right] &= & \int_{t_{\text{initial}}}^{t_{\text{final}}}\Phi\left(\bar{\sigma}\left(t\right)\right)dt\,,
%\eea
\bea
\label{sigmaBarP}
&&P\left[\bar{\sigma}\left(t\right)\right]\sim e^{-s\left[\bar{\sigma}\left(t\right)\right]/\tau_{c}} \, ,\\
\label{sigmaBarS}
&&s\left[\bar{\sigma}\left(t\right)\right]=\int_{t_{\text{initial}}}^{t_{\text{final}}}\Phi\left(\bar{\sigma}\left(t\right)\right)dt
\, , \eea
which is the continuum analog of \eqref{JPDFsigmabar}.
Eqs.~\eqref{sigmaBarP} and \eqref{sigmaBarS} provide the basis for a
coarse-grained, path-integral effective description of the telegraphic
noise, which is valid at timescales much longer than $\tau_c$.  It is
worth noting that, since the rate function $\Phi$ is parabolic at
small argument,
\be \Phi\left(z\right)\simeq z^{2}/2, \qquad |z|\ll1,\ee
it follows that at small noise amplitudes,
$\left|\bar{\sigma}\right|\ll1$, Eq.~\eqref{sigmaBarS} reduces to the
Wiener action
$$\frac{1}{2}\int_{t_{\text{initial}}}^{t_{\text{final}}}\bar{\sigma}\left(t\right)^{2}dt\,,$$
that describes white noise. However, signatures of the activity of the
noise will become noticeable for $\bar{\sigma}\sim1$.

We now return to our model given by
Eq.~\eqref{LangevinDimensionless}. At fixed $\gamma$ and in the limit
$\tau_c \to 0$, the characteristic timescales of the left hand side of
the equation (describing a deterministic damped harmonic oscillator),
which are of order unity, become much longer than $\tau_c$. Therefore,
one can replace the telegraphic noise $\sigma\left(t\right)$ by its
coarse-grained analog, i.e., one can consider the equation \be
\label{LangevinCoarseGrained}
\ddot{x}+\gamma\dot{x}+x=\bar{\sigma}\left(t\right) \, ,
\ee
which, over timescales much larger than $\tau_c$, gives a correct
effective description of the process defined by
\eqref{LangevinDimensionless}.

\subsection{Optimal fluctuation method (OFM)}

We now treat Eq.~\eqref{LangevinCoarseGrained} as the starting point
for an application of the OFM (which is sometimes referred to as
``weak-noise theory'', or as ``geometrical optics of diffusion'' in
the context of Brownian motion, and is also related to the
``macroscopic fluctuation theory'' of lattice gases). The OFM yields
the (approximate) probability of the large deviation by finding the
optimal (i.e., most likely) history of the system conditioned on the
occurrence of the rare event \cite{Onsager,MSR,
  Freidlin,Dykman,Graham, Falkovich01, GF,EK04, Ikeda2015,Grafke15,
  bertini2015, Holcman, SmithMeerson2019, MeersonSmith2019,
  AiryDistribution20}.  This optimal history is in certain contexts
referred to as an ``instanton''.
%The optimal history, an intriguing observable quantity in its own right, is obtained by using a variant of the dissipative WKB approximation, which can be cast into Hamiltonian equations of underlying dynamics of the model.

At the heart of the OFM lies a saddle-point evaluation of the path
integral representation of the process described by
Eq.~\eqref{LangevinCoarseGrained}, which exploits the small parameter
$\tau_c \ll 1$.  The usual first step of the OFM would be to plug
Eq.~\eqref{LangevinCoarseGrained} into \eqref{sigmaBarP}, in order to
obtain an expression for the probability to observe some
(coarse-grained) trajectory $x(t)$:
\bea
\label{Pxt}
P\left[x\left(t\right)\right] &\sim& e^{-\tilde{s}\left[x\left(t\right)\right]/\tau_{c}} \, ,\\
\label{sxt}
\tilde{s}\left[x\left(t\right)\right] &= & \int\Phi\left(\ddot{x}+\gamma\dot{x}+x\right)dt \, ,
\eea
up to a Jacobian whose contribution is subleading in the limit $\tau_c
\to 0$.  This leads to a minimization problem for the action
\eqref{sxt} over trajectories $x(t)$, which is to be solved subject to
boundary conditions and/or constraints that depend on the particular
problem that one is interested in solving.  We are after the
steady-state distribution $P_{\text{st}}\left(X\right)$, for which it
is appropriate to consider trajectories $x(t)$ on the time interval
$-\infty < t < 0$, under the constraints
\bea
x\left(t\to-\infty\right)&=&0 \, ,\\ x\left(t=0\right)&=&X \, .  \eea
Once the optimal trajectory -- the minimizer of \eqref{sxt} under
these constraints -- is found, $P_{\text{st}}\left(X\right)$ is
calculated by evaluating the action $s\left[x\left(t\right)\right]$ on
the optimal trajectory, leading to the scaling behavior
\eqref{PstScaling} reported above.

\begin{figure*}[ht]
\includegraphics[width=0.48\linewidth,clip=]{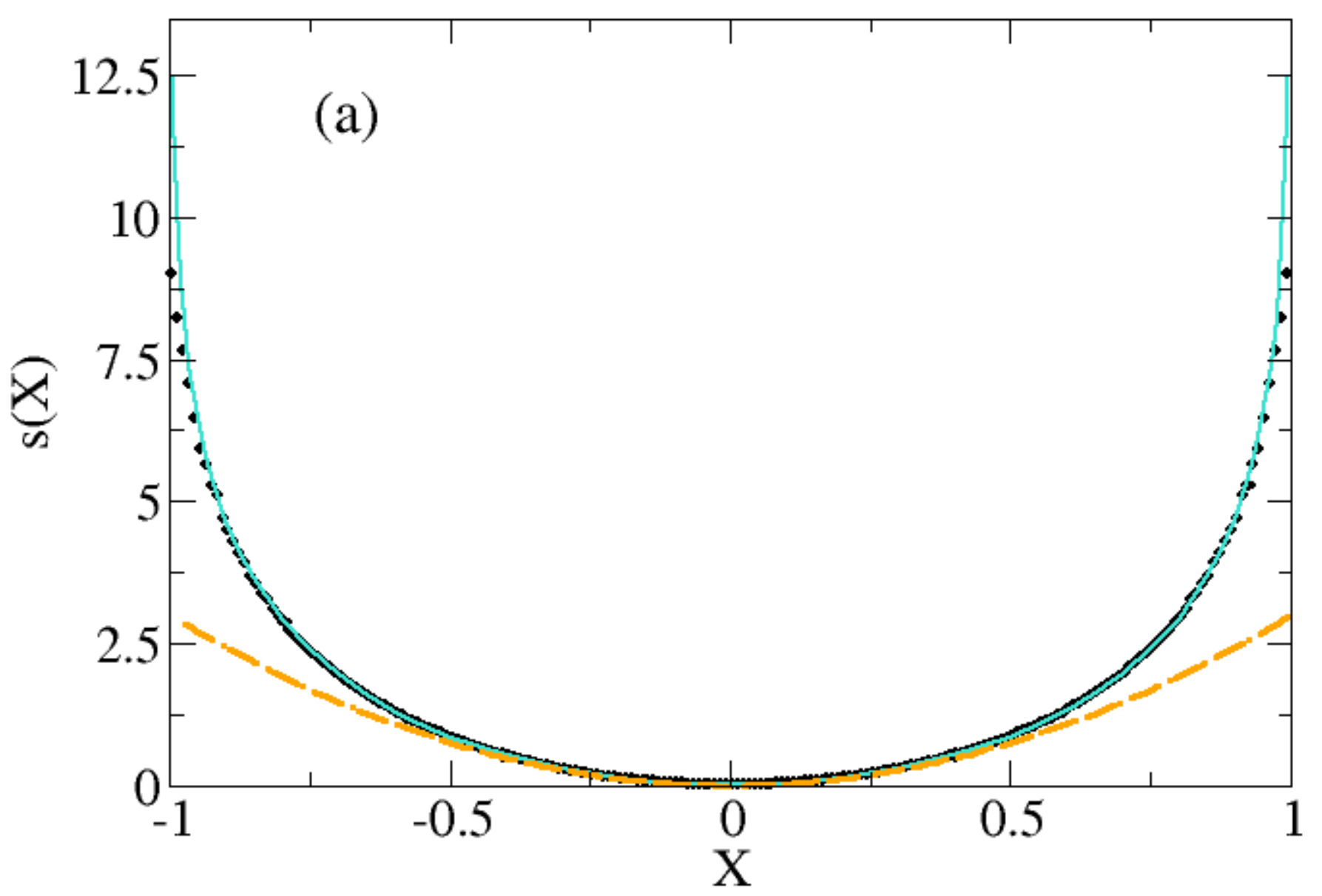}
\hspace{1mm}
\includegraphics[width=0.46\linewidth,clip=]{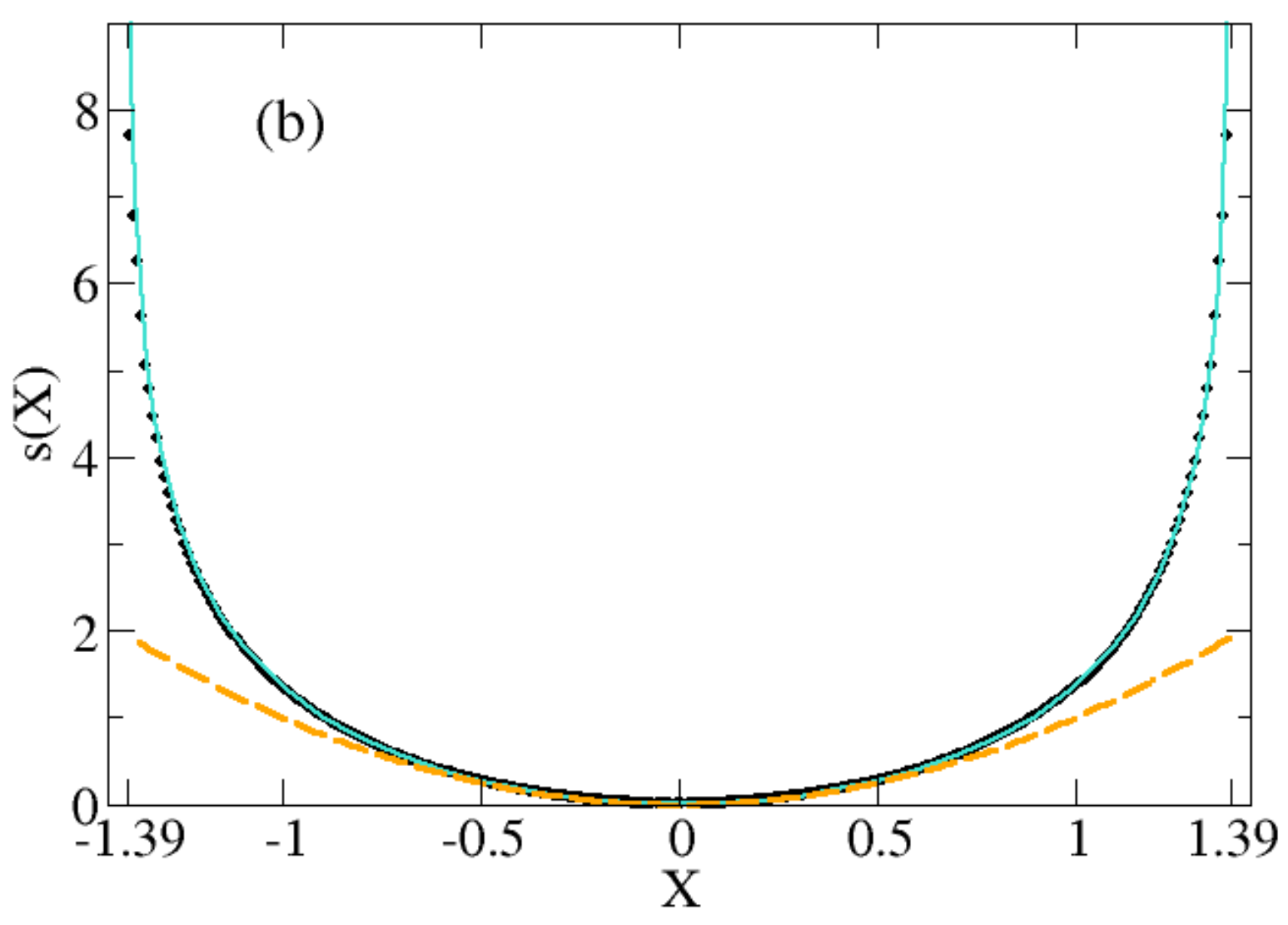}
\caption{Solid curves: The LDFs $s(X)$ for $\gamma = 3$ (a) and
  $\gamma = 1$ (b), corresponding to the overdamped ($\gamma>2$) and underdamped ($\gamma<2$)
  cases, respectively.  The most pronounced difference between
    the overdamped and underdamped regimes is the fact that
      $s(X)$ diverges at $\pm X_0$, where $X_0=1$ for $\gamma>2$ and
      $X_0>1$ for $\gamma<2$,
    %that the support $\left[-X_{0},X_{0}\right]$ of the distribution
    %changes (as described by the singularity of $X_0$ as a function
    %of $\gamma$
    see Fig.~\ref{fig:X0ofgamma} below.  Dashed curves:
    The approximation $s\left(X\ll1\right)\simeq\gamma X^{2}$,
    corresponding to the description of the steady state through a
    Boltzmann distribution with an effective temperature. Symbols:
    Results of computer simulations.
%The behavior at $X \simeq X_0$ is plotted in
  %Fig.~\ref{fig:sofXgamma3and1edge} below.}
  }
\label{fig:sofXgamma3and1}
\end{figure*}

However, the method outlined above leads to a technical difficulty in
the present case, because the Euler-Lagrange equation associated with
the functional \eqref{sxt} is a fourth-order nonlinear differential
equation, due to the term with second derivative, $\ddot{x}$, in the
functional.  In order to bypass this difficulty, we use a shortcut
which significantly simplifies the solution. The shortcut involves
reformulating the problem as an optimization problem for the
coarse-grained noise $\bar{\sigma}(t)$. The functional that is to be
minimized, \be
\label{sofSigma}
s\left[\bar{\sigma}\left(t\right)\right]=\int_{-\infty}^{0}\Phi\left(\bar{\sigma}\left(t\right)\right)dt
\, , \ee
is a lot simpler because it contains no time derivatives.  However,
the constraint $x\left(t=0\right)=X$ takes a more complicated form in
terms of $\bar{\sigma}$, which we now derive.
%Fortunately, the forced damped harmonic oscillator can be solved exactly. 
Solving Eq.~\eqref{LangevinCoarseGrained} for $x(t)$, we obtain
\be
\label{xtFromSigma}
x\left(t\right)=\int_{-\infty}^{t}\bar{\sigma}\left(t'\right)G\left(t,t'\right)dt' \, ,
\ee
where
\be
G\left(t,t'\right) = \theta\left(t-t'\right)\frac{e^{-a_{-}\left(t-t'\right)}-e^{-a_{+}\left(t-t'\right)}}{a_{+}-a_{-}} \, ,
\ee
is the Green's function for the forced damped harmonic oscillator,
i.e., the solution of the equation
\be
\partial_{t}^{2}G\left(t,t'\right)+\gamma\partial_{t}G\left(t,t'\right)+G\left(t,t'\right)=\delta\left(t-t'\right)
\ee
that satisfies $G\left(t,t'\right)=0$ at $t<t'$, and where
\be
\label{apmdef}
a_{\pm}=\frac{\gamma\pm\sqrt{\gamma^{2}-4}}{2} \, .
\ee
Using Eq.~\eqref{xtFromSigma}, the boundary condition
$x\left(t=0\right)=X$ becomes
\be
\label{Xconstraint}
X=\int_{-\infty}^{0}\bar{\sigma}\left(t\right)G\left(0,t\right)dt=\int_{-\infty}^{0}\bar{\sigma}\left(t\right)\frac{e^{a_{-}t}-e^{a_{+}t}}{a_{+}-a_{-}}dt \, .
\ee

This reformulation of the problem as an optimization problem for
$\bar{\sigma}$ makes its solution very easy. We take the constraint
\eqref{Xconstraint} into account via a Lagrange multiplier, i.e., we
minimize the modified action
\bea
\label{slambdadef}
&&s_{\lambda}\left[\bar{\sigma}\left(t\right)\right] = s\left[\bar{\sigma}\left(t\right)\right]-\lambda X \nn\\
&&\qquad=\int_{-\infty}^{0}\left[\Phi\left(\bar{\sigma}\left(t\right)\right)-\lambda\bar{\sigma}\left(t\right)\frac{e^{a_{-}t}-e^{a_{+}t}}{a_{+}-a_{-}}\right]dt\,.
\eea
The minimization becomes trivial because the functional does not
contain any derivatives in time. One simply requires the derivative of
the integrand in \eqref{slambdadef} with respect to
$\bar{\sigma}\left(t\right)$ to vanish, yielding the optimal
realization of the coarse-grained noise
\be
\label{Psol}
\bar{\sigma}\left(t\right)=\left(\Phi'\right)^{-1}\left(\lambda\frac{e^{a_{-}t}-e^{a_{+}t}}{a_{+}-a_{-}}\right) \, ,
\ee
where
\be
\label{PhiPrimeInv}
\left(\Phi'\right)^{-1}\left(y\right)=\frac{y}{\sqrt{1+y^{2}}}
\ee
is the inverse function of $d\Phi/dz$ (where we recall that $\Phi$ is
given in Eq.~\eqref{Phidef}).

In order to calculate the LDF $s(X)$, we first plug
$\bar{\sigma}\left(t\right)$ from the solution \eqref{Psol} into
Eq.~\eqref{sofSigma}, yielding $s$ as a function of $\lambda$:
\bea
\label{soflambda}
&&s=\int_{-\infty}^{0}\Phi\left(\left(\Phi'\right)^{-1}\left(\lambda\frac{e^{a_{-}t}-e^{a_{+}t}}{a_{+}-a_{-}}\right)\right)dt \nn\\
&& = \int_{-\infty}^{0}\left(1-\frac{a_{+}-a_{-}}{\sqrt{\left(a_{+}-a_{-}\right)^{2}+\lambda^{2}\left(e^{a_{-}t}-e^{a_{+}t}\right)^{2}}}\right)dt \, . \nn\\
\eea
The relation between $\lambda$ and $X$ is then found by plugging
\eqref{Psol} into the constraint \eqref{Xconstraint}, yielding:
%\be
%\label{Xoflambda}
%X= \! \int_{-\infty}^{0}\frac{\lambda\left(e^{a_{-}t}-e^{a_{+}t}\right)^{2}}{\left(a_{+}-a_{-}\right)\sqrt{\left(a_{+}-a_{-}\right)^{2}+\lambda^{2}\left(e^{a_{-}t}-e^{a_{+}t}\right)^{2}}}dt  .
%\ee
\bea
\label{Xoflambda}
&&X=\int_{-\infty}^{0}\frac{e^{a_{-}t}-e^{a_{+}t}}{a_{+}-a_{-}}\left(\Phi'\right)^{-1}\left(\lambda\frac{e^{a_{-}t}-e^{a_{+}t}}{a_{+}-a_{-}}\right)dt\nn\\ &&=\!\int_{-\infty}^{0}\frac{\lambda\left(e^{a_{-}t}-e^{a_{+}t}\right)^{2}}{\left(a_{+}-a_{-}\right)\sqrt{\left(a_{+}-a_{-}\right)^{2}+\lambda^{2}\left(e^{a_{-}t}-e^{a_{+}t}\right)^{2}}}dt.\nn\\ \eea
Eqs.~\eqref{soflambda} and \eqref{Xoflambda} give the LDF $s(X)$ in a
parametric form, and constitute the main theoretical result of this
paper. This function $s(X)$ is plotted in the solid line in
Fig.~\ref{fig:sofXgamma3and1} for (a) $\gamma = 3$ and  (b) $\gamma = 1$, corresponding to the overdamped and underdamped
cases, respectively. The LDF $s(X)$ has a parabolic behavior around
$X=0$ (displayed by the dashed curves in
Fig.~\ref{fig:sofXgamma3and1}), and it diverges at $X=X_0$.
%, as described below .
The circle symbols denote the results of computer simulations which,
indeed, show a very good agreement with the analytical
predictions. The details of the computational methods are
described in Appendix \ref{app:simulations}.

 Note that the dependence of our main results on the particular
  type of noise (which we took to be dichotomous) enters only through
  the rate function $\Phi(z)$. Therefore, for other types of active
  noise, the SSD can still be found, by first calculating the
  corresponding rate function $\Phi(z)$ and then plugging into the
  first lines of Eqs.~\eqref{soflambda} and \eqref{Xoflambda} to
  obtain the LDF $s(X)$.

\begin{figure}[ht]
\includegraphics[width=0.98\linewidth,clip=]{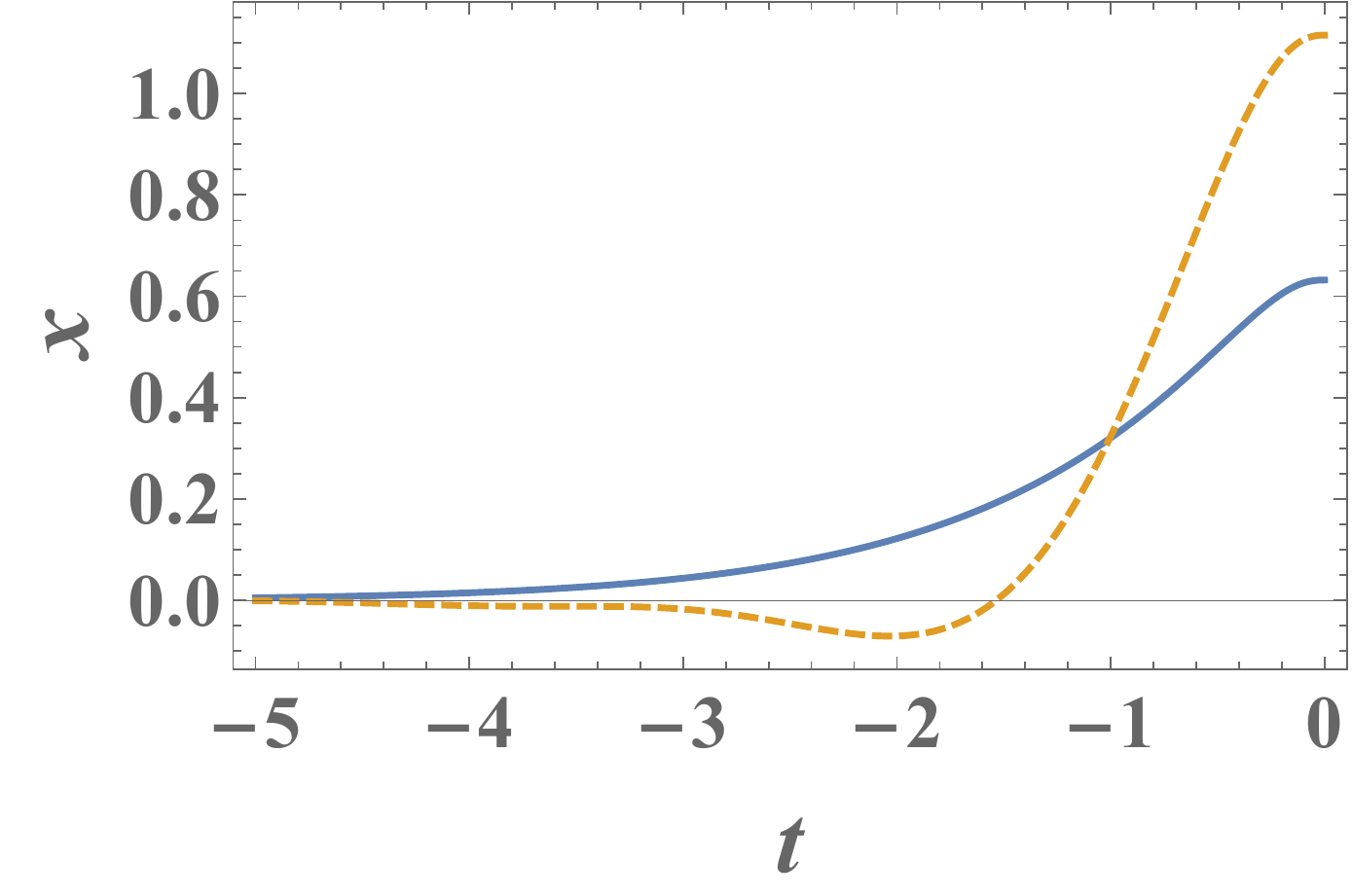}
\caption{Optimal histories of the system $x(t)$ conditioned on
  reaching position $X$ at time $0$, see Eq.~\eqref{xtsol}. Parameters
  are $\gamma = 3$ and $\lambda = 6$ for the overdamped case (solid
  line) and $\gamma = 1$ and $\lambda = 6$ for the underdamped case
  (dashed line), corresponding to $X=0.632168\dots$ and
  $X=1.1154\dots$, respectively.}
\label{fig:xoft}
\end{figure}

The OFM gives us additional useful information - the optimal
(coarse-grained) realization of the process $x(t)$, conditioned on the
constraint $x(0) = X$. This is now simply found by plugging
Eq.~\eqref{Psol} into \eqref{xtFromSigma}:
%, see Appendix \ref{app:opdfsdfsd}.
%Although unnecessary for the purpose of calculating the LDF $s(X)$, the OFM also yields the optimal coarse-grained realization $x(t)$ of the process, conditioned on $x(t=0)=X$.  This is given by plugging Eq.~\eqref{Psol} into \eqref{xtFromSigma}:
\bea
\label{xtsol}
x\left(t\right)&=&\int_{-\infty}^{t}\left(\Phi'\right)^{-1}\left(\lambda\frac{e^{a_{-}t'}-e^{a_{+}t'}}{a_{+}-a_{-}}\right)
\nn\\ &\times&
\frac{e^{-a_{-}\left(t-t'\right)}-e^{-a_{+}\left(t-t'\right)}}{a_{+}-a_{-}}dt'
\, .  \eea
Optimal paths are plotted in Fig.~\ref{fig:xoft} for the
overdamped and underdamped cases. These examples illustrate a
  general feature of the optimal trajectories: In the overdamped case
$x$ is a monotonic function of $t$, whereas in the underdamped case
$x$ oscillates with an amplitude that grows in time.
%exhibits oscillations that grow in time.
 This feature is not a consequence of the activity in the system, as it occurs also for the case of thermal noise, as will become immediately clear from the discussion of the case $X \ll 1$.

In the limit $X \ll 1$ (in which, as we will explain shortly in subsection \ref{sec:typical}, the noise can be approximated as thermal, with no signature of activity in the leading order), the optimal trajectory becomes [using
  $\left(\Phi'\right)^{-1}\left(y\ll1\right)\simeq y$ in
  \eqref{xtsol}]:
\be
x\left(t\right)\simeq\frac{\lambda}{2\left(a_{+}-a_{-}\right)\left(a_{-}+a_{+}\right)}\left(\frac{e^{a_{-}t}}{a_{-}}-\frac{e^{a_{+}t}}{a_{+}}\right)\,.
\ee
This is nothing but the time-reversed relaxation trajectory for a
particle that begins at position $X$ with zero velocity, and evolves
in time in the absence of noise, i.e., according to
$\ddot{x}+\gamma\dot{x}+x=0$. This is expected due to Onsager-Machlup
symmetry for equilibrium systems \cite{Onsager}, that follows from the
time-reversal symmetry of the dynamics in the case of thermal noise.

\subsection{Asymptotic limits}

The results given above give the exact LDF, $s(X)$, at any $X$ and any
$\gamma$. We now study the behavior of $s(X)$ in three limiting cases:
Typical fluctuations $\left|X\right|\ll1$, the behavior near the edge
$X_{0}-\left|X\right|\ll X_{0}$, and the strongly-overdamped regime
$\gamma \gg 1$.

\subsubsection{Typical fluctuations $\left|X\right|\ll1$}
\label{sec:typical}

In the regime of typical fluctuations, $\left|X\right|\ll1$, or
equivalently $\left|\lambda\right|\ll1$, Eqs.~\eqref{soflambda} and
\eqref{Xoflambda} simplify to
\be
s\left(\lambda\ll1\right)\simeq\frac{\lambda^{2}}{2\left(a_{+}-a_{-}\right)^{2}}\int_{-\infty}^{0}\left(e^{a_{-}t}-e^{a_{+}t}\right)^{2}dt=\frac{\lambda^{2}}{4\gamma}
\ee and \be X\left(\lambda\ll1\right)\simeq
\lambda\int_{-\infty}^{0}\frac{\left(e^{a_{-}t}-e^{a_{+}t}\right)^{2}}{\left(a_{+}-a_{-}\right)^{2}}dt
= \frac{\lambda}{2\gamma} \ee respectively, so $s(X)$ is explicitly
given by a parabola \be
\label{sParabola}
s\left(X\ll1\right)\simeq\gamma X^{2} \, .
\ee
Plugging Eq.~\eqref{sParabola} into \eqref{PstScaling}, one finds that
typical fluctuations follow a Gaussian distribution, which, in the
physical variables, is given by
\be
\label{PGaussian}
P_{\text{st}}\left(X\right)\sim e^{-\frac{\Gamma
    kX^{2}}{\mathcal{T}_{c}\Sigma_{0}^{2}}} \, .  \ee

We now show that Eq.~\eqref{PGaussian} coincides with the result that
one obtains by approximating the noise as white.  The autocorrelation
function of the noise is
\be \left\langle
\Sigma\left(t\right)\Sigma\left(t'\right)\right\rangle
=\Sigma_{0}^{2}e^{-2\left|t-t'\right|/\mathcal{T}_{c}} \, ,
\ee
which satisfies \be \int_{-\infty}^{\infty}\left\langle
\Sigma\left(t\right)\Sigma\left(t'\right)\right\rangle
dt'=\Sigma_{0}^{2}\mathcal{T}_{c} \, .  \ee As discussed at the
introduction, the telegraphic noise does not become white in the limit
$\mathcal{T}_{c} \to 0$ at \emph{fixed} $\Sigma_0$
\cite{footnote:whiteNoise}. However, in the typical fluctuations
regime $|X|\ll1$, the white-noise approximation works well: The
telegraphic noise can be approximated by an effective white (Gaussian)
noise $\xi\left(t\right)$ with zero mean and correlation function
$\left\langle \xi\left(t\right)\xi\left(t'\right)\right\rangle
=\Sigma_{0}^{2}\mathcal{T}_{c}\delta\left(t-t'\right)$, which would
physically describe a thermal noise with an effective temperature
$T_{\text{eff}}=\Sigma_{0}^{2}\mathcal{T}_{c}/\left(2k_{B}\Gamma\right)$.
This approximation yields a Boltzmann steady-state distribution, \be
P_{\text{st}}\left(X\right)\sim
e^{-\frac{V\left(X\right)}{k_{B}T_{\text{eff}}}}, \ee which, after
plugging in the value of $T_{\text{eff}}$ as well as
$V_{k}\left(x\right)=kx^{2}/2$, indeed coincides with
\eqref{PGaussian}.

Eq.~\eqref{sParabola} can be seen to correctly describe the
small-$|X|$ behavior of the exact $s(X)$ in
Fig.~\ref{fig:sofXgamma3and1} for $\gamma = 1$ and $\gamma=3$. As is
seen in the figure, $s(X)$ lies above its parabolic approximation
\eqref{sParabola}, leading to probabilities
$P_{\text{st}}\left(X\right)$ that are smaller. Physically, this means
that the boundedness of the noise makes it less likely to reach large
$X$'s, when compared to the white-noise case. This effect is very
small at small $X$, and becomes more important as $X$ is increased,
while at $X > X_0$ it dominates and leads to
$P_{\text{st}}\left(X\right) = 0$.

\begin{figure*}[ht]
\includegraphics[width=0.47\linewidth,clip=]{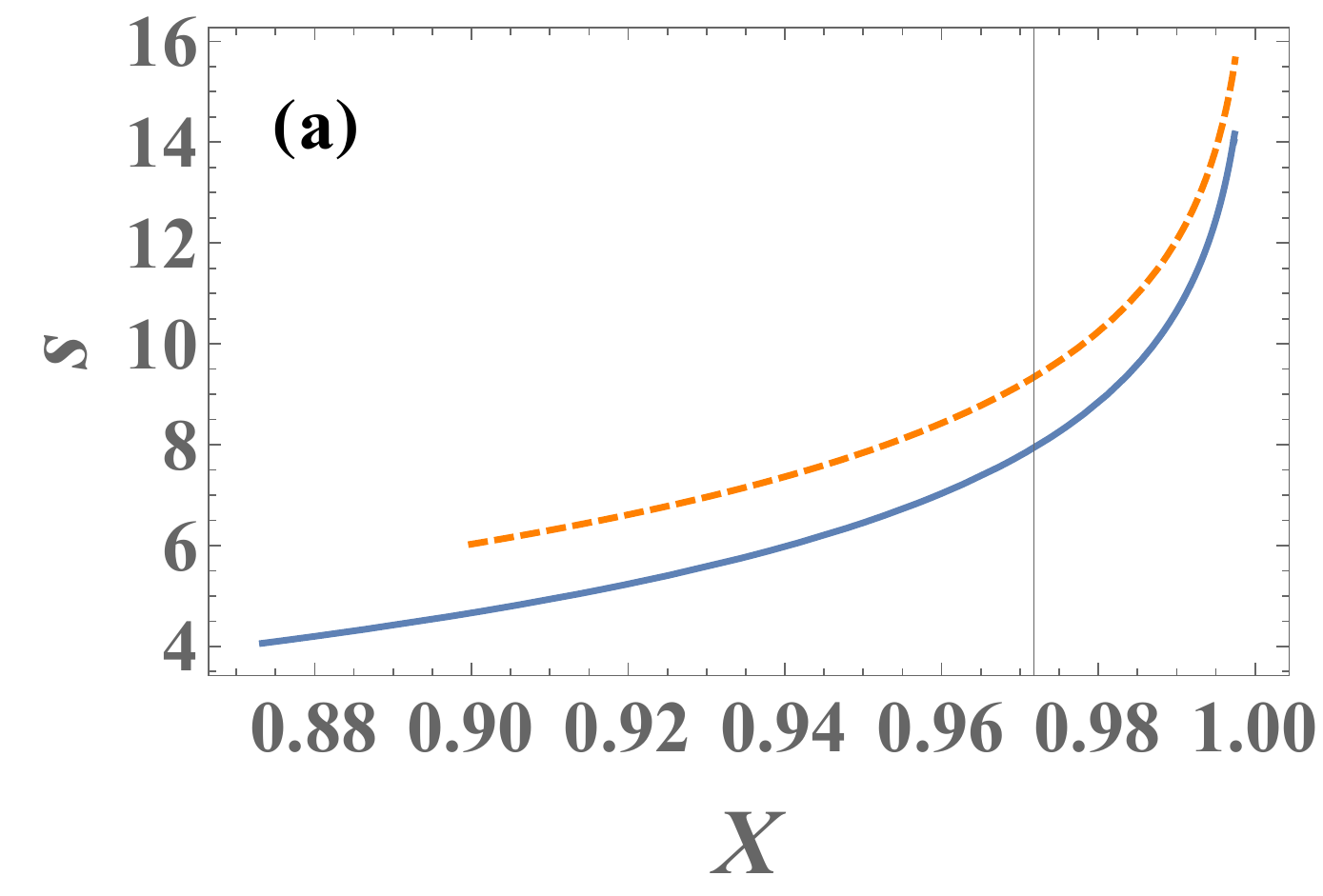}
\hspace{1mm}
\includegraphics[width=0.47\linewidth,clip=]{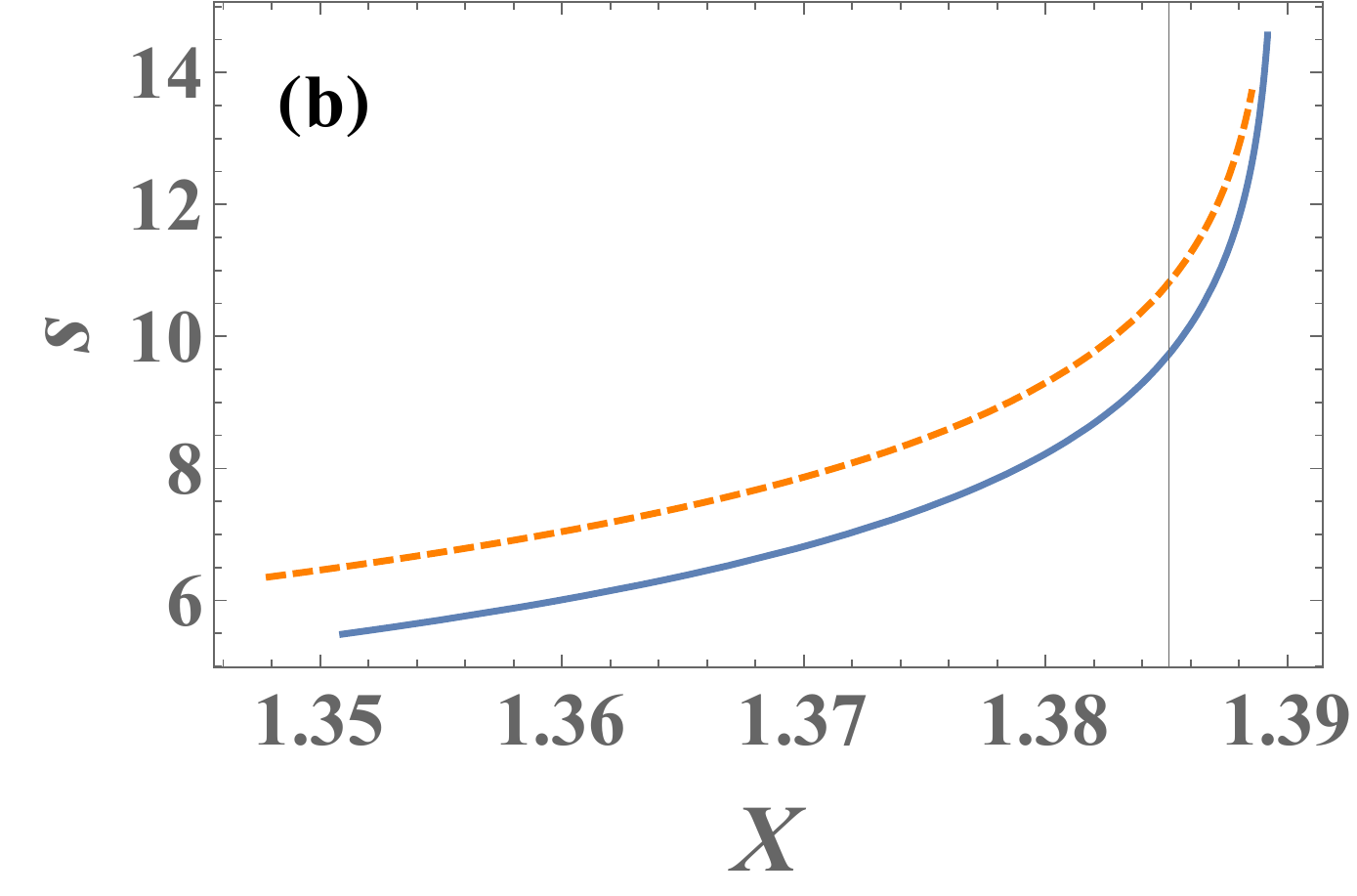}
\caption{Solid lines: The LDF $s(X)$ for $\gamma = 3$ (a) and
  $\gamma = 1$ (b), as in Fig.~\ref{fig:sofXgamma3and1}, at $X \simeq
  X_0$. Dashed lines: the approximations near the edges, given by
  Eqs.~\eqref{slargelambdaOD} and \eqref{slargelambdaUD},
  respectively. In (b), $X_0 =
  \coth\left(\frac{\pi}{2\sqrt{3}}\right)=1.38958\dots$.}
\label{fig:sofXgamma3and1edge}
\end{figure*}

\subsubsection{Near the edges of the support, $X_{0}-\left|X\right|\ll X_{0}$}

Let us now study the opposite limit, in which the particle is very
close to the edges, $X_{0}-X \ll X_{0}$, which corresponds to
$\lambda\to\infty$.  In order to do so, we first analyze the behavior
of $\bar{\sigma}\left(t\right)$ from Eq.~\eqref{Psol}, at $\lambda \gg
1$.  We consider first the overdamped case, $\gamma > 2$.  At
sufficiently early times, the exponential terms are negligible, so
$\bar{\sigma}\left(t\right)$ vanishes in the leading order.  From time
\be
\label{tlambdadef}
t_{\lambda} =
\frac{1}{a_{-}}\ln\left(\frac{a_{+}-a_{-}}{\lambda}\right)
\ee
onwards, the term $\lambda e^{a_{-}t}$ becomes very large and
dominates. As a result, using
$\left(\Phi'\right)^{-1}\left(y\gg1\right)\simeq1$ we find
$\bar{\sigma}\left(t\right) \simeq 1$.  Finally, very close to $t=0$,
one has $e^{a_{-}t}-e^{a_{+}t} \simeq 0$ ultimately leading to
$\bar{\sigma}\left(t=0\right)=0$.  To summarize, \be
\label{Papprox}
\bar{\sigma}\left(t\right)\simeq\begin{cases}
0, & t<t_{\lambda}\text{ or }t\simeq0,\\[1mm]
1, & t_{\lambda}<t<0.
\end{cases}
\ee
We verified the validity of the approximation \eqref{Papprox} in
Appendix \ref{app:nearEdges}.

The action is very easy to calculate in this limit. From
Eq.~\eqref{sofSigma}, one simply obtains \be
\label{sOD1}
s\left(\lambda\gg1\right)\simeq-t_{\lambda}\simeq\frac{\ln\lambda}{a_{-}}\,,
\ee
where we neglected the term $a_{+}-a_{-}$ in \eqref{tlambdadef} in
order to avoid excess of accuracy.  The relation between $X$ and
$\lambda$ also simplifies in this limit. Plugging Eq.~\eqref{Papprox}
into \eqref{Xconstraint} (replacing $\bar{\sigma} \to P$,) we obtain
\be
\label{XOD1}
 X\simeq\int_{t_{\lambda}}^{0}\frac{e^{a_{-}t}-e^{a_{+}t}}{a_{+}-a_{-}}dt\simeq1-\frac{1}{a_{-}\lambda} \, .
 \ee 
%  In order to obtain the rate function $s(X)$, it is convenient again to use the shortcut \eqref{dsdlambda}. We have
%\be
%\frac{dX}{d\lambda}=\frac{1}{\lambda}\frac{ds}{d\lambda}\simeq\frac{1}{a_{-}\lambda^{2}} \, ,
%\ee 
%which we immediately integrate to obtain
%\be
%X\simeq 1-\frac{1}{a_{-}\lambda} \, ,
%\ee
%where we used $X_0 = 1$ in order to determine the integration constant.
This result, in particular, implies that $X_0 = 1$ for the overdamped
case, corresponding to the particle's mechanical equilibria for the
situation in which the noise is $\sigma = 1$.  Putting
Eqs.~\eqref{sOD1} and \eqref{XOD1} together, we obtain the asymptotic
behavior
\be
\label{slargelambdaOD}
s\left(X\right)\simeq-\frac{1}{a_{-}}\ln\left(1-X\right),\quad1-X\ll1 \, ,
\ee
%$\mathcal{I}\left(t\right)$ is the optimal realization of the coarse-grained noise term, $\bar{\sigma}(t)$. 
%Therefore, 
see Fig.~\ref{fig:sofXgamma3and1edge} (a).  The physical picture in
this limit is quite simple: The noise $\bar{\sigma}\left(t\right)$
acts as a constant force, pushing the particle towards the position
$X$, for duration $\left|t_{\lambda}\right|$.

Let us now consider the underdamped case, $\gamma < 2$.  In this case,
Eq.~\eqref{Psol} takes the form
\be
\label{PsolUnderdamped}
\bar{\sigma}\left(t\right)=\left(\Phi'\right)^{-1}\left(-\frac{2\lambda}{\sqrt{4-\gamma^{2}}}e^{\gamma
  t/2}\sin\left(\frac{\sqrt{4-\gamma^{2}}}{2}t\right)\right)\,.
\ee
Eq.~\eqref{PsolUnderdamped} is exact. Let us now analyze its behavior
in the limit $\lambda \gg 1$.  At sufficiently early times, the
exponential decay again leads to $\bar{\sigma}\left(t\right) \simeq
0$.  From time $t_{\lambda} = -2\left(\ln\lambda\right)/\gamma$
%\be
%\label{tlambdaUDdef}
%t_{\lambda}\simeq-2\left(\ln\lambda\right)/\gamma
%\ee
onwards, the exponential term becomes very large. As a result, the
argument of $\left(\Phi'\right)^{-1}$ becomes much larger than 1 in
absolute value, except for short temporal boundary layers in which the
term with $\sin$ vanishes, signaling a sign change of
$\bar{\sigma}\left(t\right)$.  Using 
\be
\left(\Phi'\right)^{-1}\left(y\right)\simeq\text{sgn}\left(y\right),\qquad\left|y\right|\gg1,
\ee
we thus obtain
\be
\label{PapproxUD}
\bar{\sigma}\left(t\right)\simeq\begin{cases} 0, &
t<t_{\lambda},\\[2mm]
-\text{sgn}\left(\sin\left(\frac{\sqrt{4-\gamma^{2}}}{2}t\right)\right),
& t_{\lambda}<t<0
\end{cases}
\ee
(see Appendix \ref{app:nearEdges} for a check of this approximation).
This leads, using
Eq.~\eqref{sofSigma}, to
\be
\label{soflambdaUD}
s\left(\lambda\gg1\right)\simeq-t_{\lambda}=\frac{2\ln\lambda}{\gamma}\,.
\ee

One way to find the connection between $X$ and $\lambda$ is to analyze
the expression \eqref{Xoflambda} in the limit $\lambda \gg 1$.
However, this turns out to be rather difficult from a technical point
of view. Instead, we will employ a useful shortcut that bypasses this
calculation.  The shortcut makes use of the relation $ds/dX=\lambda$,
a property that follows from the fact that $X$ and $\lambda$ are
conjugate variables, see \textit{e.g.} Ref. \cite{Vivoetal}.  Using
this property together with the chain rule, we obtain \be
\label{dsdlambda}
\frac{ds}{d\lambda}=\frac{ds}{dX}\frac{dX}{d\lambda}=\lambda\frac{dX}{d\lambda}
\, .  \ee Using Eq.~\eqref{soflambdaUD}, we find \be
\frac{dX}{d\lambda}=\frac{1}{\lambda}\frac{ds}{d\lambda}\simeq\frac{2}{\gamma\lambda^{2}}\,,
\ee which we immediately integrate to obtain \be X\simeq
X_{0}-\frac{2}{\gamma\lambda}\,, \ee where the integration constant
$X_0 $ will be determined shortly.  Putting these results together, we
obtain the asymptotic behavior
%
%A similar calculation to the one that we performed in the overdamped case then yields
\be
\label{slargelambdaUD}
s\left(X\right)\simeq-\frac{2}{\gamma}\ln\left(X_{0}-X\right),\qquad X_{0}-X\ll X_0\,,
\ee
see Fig.~\ref{fig:sofXgamma3and1edge} (b).

Let us now calculate $X_0$.  Plugging Eq.~\eqref{PapproxUD} into
\eqref{Xconstraint},
%(replacing $\bar{\sigma} \to P$,)
we obtain \bea
X&=&-\frac{2}{\sqrt{4-\gamma^{2}}}\int_{-\infty}^{0}\bar{\sigma}\left(t\right)e^{\gamma
  t/2}\sin\left(\frac{\sqrt{4-\gamma^{2}}}{2}t\right)dt\simeq\nn\\ &\simeq&\frac{2}{\sqrt{4-\gamma^{2}}}\int_{t_{\lambda}}^{0}e^{\gamma
  t/2}\left|\sin\left(\frac{\sqrt{4-\gamma^{2}}}{2}t\right)\right|dt
\, .  \eea In order to calculate $X_0$ as a function of $\gamma$, we
simply take the limit $\lambda \to \infty$, in which one has
$t_\lambda \to -\infty$, so \bea
\label{X0sol}
X_{0}&=&\frac{2}{\sqrt{4-\gamma^{2}}}\int_{-\infty}^{0}e^{\gamma
  t/2}\left|\sin\left(\frac{\sqrt{4-\gamma^{2}}}{2}t\right)\right|dt
\nn\\ &=& \coth\left(\frac{\pi\gamma}{2\sqrt{4-\gamma^{2}}}\right) \,
.  \eea
The details of the solution of the integral in Eq.~\eqref{X0sol} are
given in Appendix \ref{app:X0}. In Fig.~\ref{fig:X0ofgamma}, $X_0$ is
plotted as a function of $\gamma$.  The particle cannot reach the
region $X > X_0$, i.e., $P_{\text{st}}\left(X\right)$ vanishes
there. The divergence of the LDF at $X \to X_0$, as given by
Eqs.~\eqref{slargelambdaOD} and \eqref{slargelambdaUD}, describes the
manner in which $P_{\text{st}}\left(X\right)$ vanishes as $X \to X_0$.

 \begin{figure}[ht]
\includegraphics[width=0.98\linewidth,clip=]{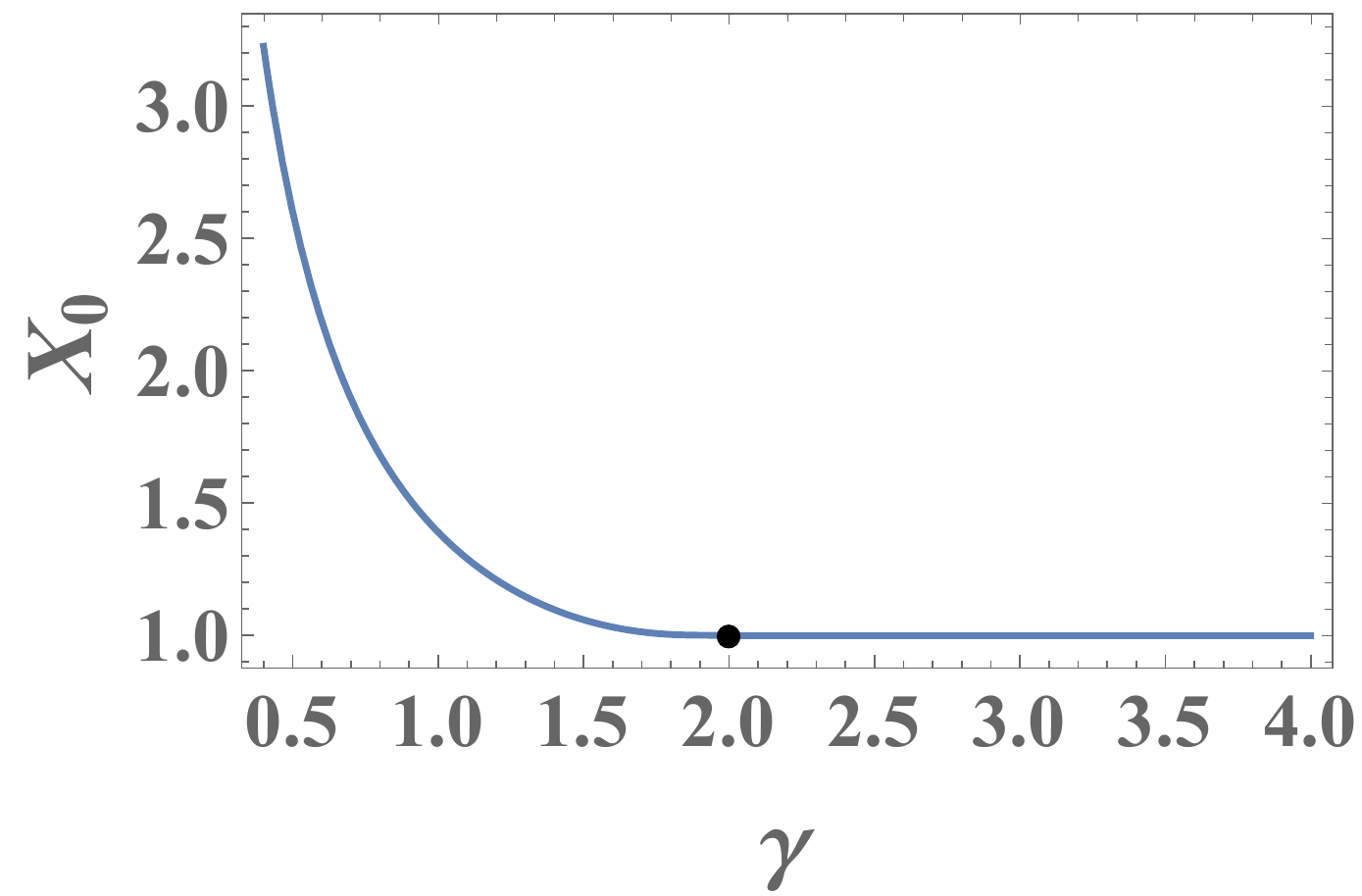}
\caption{The maximal possible position of the particle, $X_0$, as a
  function of the damping coefficient $\gamma$, which is given by
  Eq.~\eqref{X0sol} in the underdamped case $\gamma < 2$ and by $X_0 =
  1$ for $\gamma > 2$.}
\label{fig:X0ofgamma}
\end{figure}

\subsubsection{Strongly-overdamped regime $\gamma \gg 1$}

Here we study the strongly-overdamped regime, $\gamma \gg 1$, or $m
\to 0$ in the physical variables. In this limit, the motion described
by Eq.~\eqref{LangevinDimensional} can be interpreted as that of a
run-and-tumble particle confined in a harmonic potential
\cite{Kardar15, Maes2018, Dhar_2019, Basu20}.  As we mentioned in the
Introduction, in this limit the SSD is known exactly for telegraphic
noise with any trapping potential $V(x)$
\cite{VBH84, q-optics1,q-optics2,q-optics3,q-optics4,colored, Kardar15, Dhar_2019,
  Basu20}.  Up to a normalization factor, the SSD is given, in the
physical variables, by
%\be
%\label{eq:Pst_2st}
%P_{\text{st}}\left(X\right)\propto\frac{1}{v_{0}^{2}-f^{2}(x)}\exp\left[\frac{2}{\mathcal{T}_{c}}\int_{0}^{X}dy\frac{f(y)}{v_{0}^{2}-f^{2}(y)}\right]
%\ee
%where $f\left(x\right)=-V'\left(x\right) / \Gamma$ is the force acting on the particle, and $v_0 = \Sigma_0 / \Gamma$.
\be
P_{\text{st}}\left(X\right)\propto\frac{\Gamma^{2}}{\Sigma_{0}^{2}-F^{2}(X)}\exp\left[\frac{2\Gamma}{\mathcal{T}_{c}}\int_{0}^{X}dy\frac{F(y)}{\Sigma_{0}^{2}-F^{2}(y)}\right]\,,
\ee where $F\left(x\right)=-V'\left(x\right)$ is the force that acts
on the particle. In the case of the harmonic potential $V(x) = kx^2
/2$, this becomes \cite{Dhar_2019, TC08} \be
\label{PstExactHO}
%P_{\text{st}}\left(X\right)\propto \frac{\Gamma^{2}}{\Sigma_{0}^{2}-k^{2}X^{2}}\exp\left[\frac{\Gamma}{k\mathcal{T}_{c}}\ln\left(\Sigma_{0}^{2}-k^{2}X^{2}\right)\right] \, .
P_{\text{st}}\left(X\right)\propto\exp\left[\left(\frac{\Gamma}{k\mathcal{T}_{c}}-1\right)\ln\left(\Sigma_{0}^{2}-k^{2}X^{2}\right)\right]
\, .  \ee
In the limit $\mathcal{T}_c \to 0$, the term `$-1$' in the exponential
in Eq.~\eqref{PstExactHO} is negligible compared to the term
$\Gamma/k\mathcal{T}_{c}$, and one recovers our Eq.~\eqref{PstScaling}
with the LDF \be
\label{sOverDamped}
s(X) = - \gamma\ln\left(1-X^{2}\right)
\ee
(in the rescaled variables).

\begin{figure}[ht]
\includegraphics[width=0.98\linewidth,clip=]{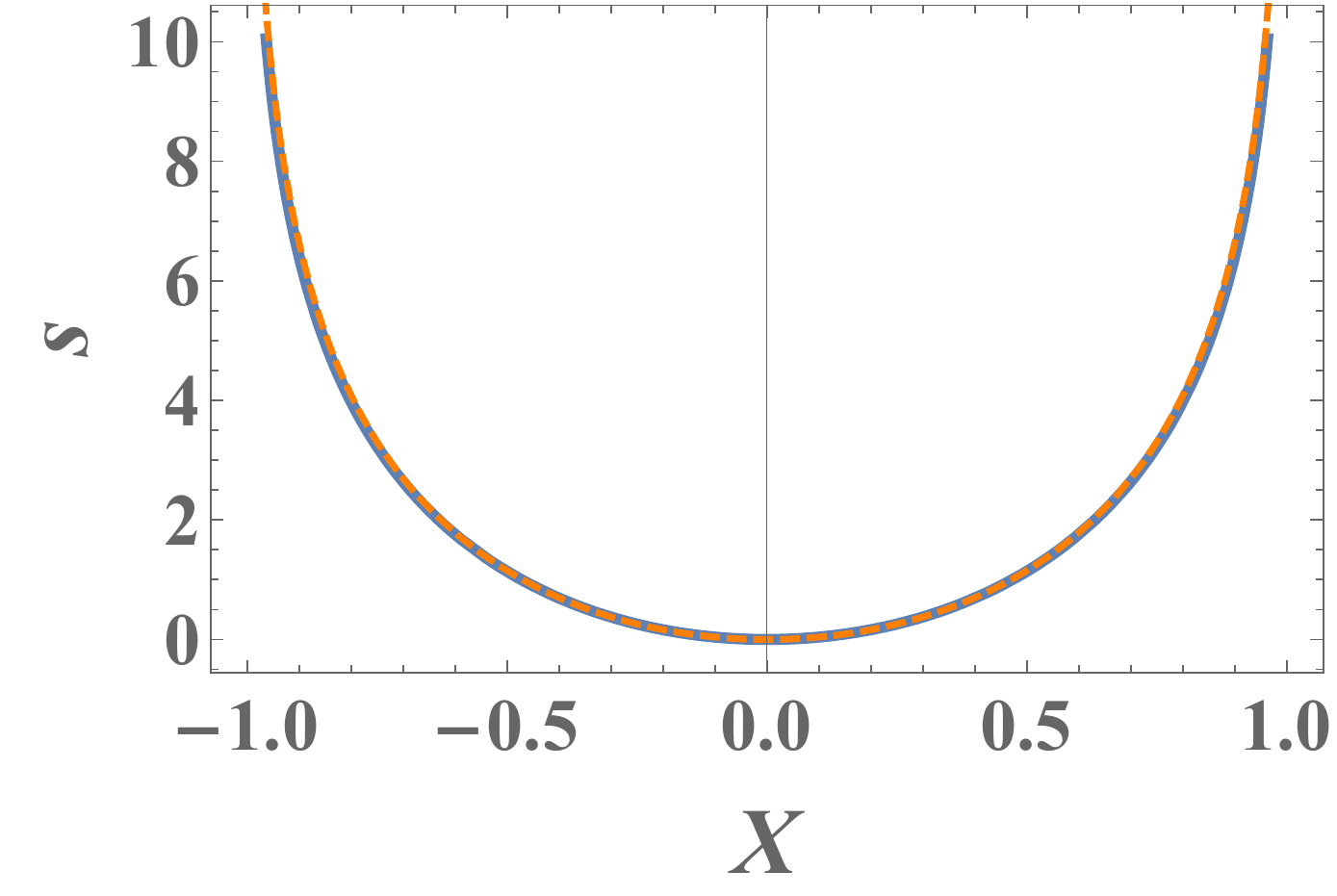}
\caption{Solid line: The exact LDF  $s(X)$ for $\gamma =
  4$. This value of $\gamma$ turns out to be sufficiently large for
  the strongly-overdamped approximation \eqref{sOverDamped} (dashed
  line) to work very well, as is clearly seen in the figure.}
\label{fig:sofXgammat}
\end{figure}

Let us show that the LDF given by Eqs.~\eqref{soflambda} and
\eqref{Xoflambda} indeed recovers this result when $\gamma \gg
1$.  In this limit, Eq.~\eqref{apmdef} becomes \be a_+ \simeq
\gamma, \qquad a_- \simeq \frac{1}{\gamma} \, , \ee so $a_+ \gg
a_-$. As a result, the two exponential terms in Eq.~\eqref{soflambda}
decay over very different timescales, so we can safely discard the
faster-decaying one, leading to \be
\label{soflambdaOD1}
s \simeq
\int_{-\infty}^{0}\left(1-\frac{\gamma}{\sqrt{\gamma^{2}+\lambda^{2}e^{2t/\gamma}}}\right)dt
\, .  \ee In Appendix \ref{Appendix:OD} we solve this integral, and
then use the shortcut \eqref{dsdlambda} yielding explicit expressions
for $s$ and $X$ as functions of $\lambda$, from which we indeed
recover Eq.~\eqref{sOverDamped}.  In Fig.~\ref{fig:sofXgammat}, the
exact $s(X)$ is plotted together with the approximation
\eqref{sOverDamped} for $\gamma=4$. The two plots are
indistinguishable within the scale of the figure.

\section{Conclusion}
\label{sec:conclusion}

In this paper, we calculated the SSD (and from it, the MFPT) of a
damped particle which experiences an external harmonic force, and is
driven by telegraphic (dichotomous) noise, in the limit where the
switching rate of the noise is very large.  In the
typical-fluctuations regime, the noise can be approximated as white,
leading to a Boltzmann distribution for the SSD and an Arrhenius law
for the MFPT, with an effective temperature that is easy to calculate
from the correlation function of the noise.  However, in the
large-deviations regime, the effective thermal description of the
system breaks down.  We found that both the SSD and the MFPT follow
large-deviation principles, Eqs.~\eqref{PstScaling} and
\eqref{SofXdef} respectively. We proceeded to calculate the
large-deviation function $s(X)$ analytically, see
Eqs.~\eqref{soflambda} and \eqref{Xoflambda}, showing very good
agreement with Langevin Dynamics simulations (see also Appendix
  \ref{app:simulations}).
%We also extended our results to other types of active noise.

We found that, at a coarse-grained temporal scale, the dominant
contribution to the probability for observing the particle at some
position $X$ comes from an optimal trajectory $x(t)$ [and a
corresponding optimal coarse-grained noise realization
$\bar{\sigma}\left(t\right)$].
%that we calculated exactly
We observed a qualitative difference in the behavior between the
overdamped and underdamped regimes: In the former, the noise
$\bar{\sigma}\left(t\right)$ pushes $x(t)$ monotonically towards $X$,
whereas in the latter, $x(t)$ oscillates with a growing amplitude,
while $\bar{\sigma}\left(t\right)$ changes sign every
half-oscillation, increasing the particle's energy until it reaches
$X$.  In both cases, the particle is confined in space
$x\left(t\right)<X_{0}$, with $X_0 = 1$ in the overdamped case, but
$X_0 > 1$ in the underdamped case, see Eq.~\eqref{X0sol}.  Notably,
our coarse-graining procedure maintains the nonequlibrium (active)
character of the noise, cf. \cite{HG21}.

It is worth mentioning the recent work \cite{AgranovBunin21}, in which
the stochastic dynamics of two populations was studied, and a
separation of timescales was exploited in order to employ a `hybrid'
DV-OFM approach that is similar in spirit to ours. In the current work
we point out that this method can be applied quite generally, and
demonstrate that it is especially useful in systems with non-Gaussian
noise with a short correlation time.

Our main result for $s(X)$ can be straightforwardly extended to other
types of active noise whose associated rate function $\Phi(z)$ can be
found, by simply plugging $\Phi(z)$ into the first lines of
Eqs.~\eqref{soflambda} and \eqref{Xoflambda}.
A second immediate extension is to a noise term $\xi(t)$ in
Eq.~\eqref{Langevin} (with a harmonic potential) that is given by the
sum of two (or more) statistically-independent noises, e.g., the sum
of active and thermal noises \cite{BenIsaac15, Wexler20, Woillez20,
  GP21, Tucci22}.  Due to the linearity of the equation, one finds
that the SSD is given (exactly) by the convolution of the SSDs that
correspond to each of the two noise terms separately \cite{Tucci22, Frydel22, 
  SLMS22}.  Finally, it would be interesting to extend our results to
anharmonic potentials \cite{Wexler20,led20} higher spatial dimensions
\cite{ABP2018, Malakar20, Basu20, SBS20, MB20, SBS21, Frydel22, SBS22,
  SLMS22}, and to interacting run-and-tumble particles \cite{LMS21}.

\subsection*{Acknowledgments}
NRS acknowledges helpful discussions with Golan Bel, Tal Agranov and
Baruch Meerson and is grateful to Pierre Le Doussal, Satya Majumdar
and Grégory Schehr for a collaboration on related topics.   We
  are grateful to Yariv Kafri for a helpful correspondence and for
  pointing out several references.

\appendix

\medskip

\section{Relation between the FPT distribution and the SSD}
\label{app:MFPTandSSD}

To reach the relation 
\be
\label{MFPTandSSD}
\left\langle t_{\text{esc}}\left(X\right)\right\rangle \sim1/P_{\text{st}}\left(X\right)
\ee
between the MFPT and the SSD that is given in Eq.~\eqref{SofXdef} of
the main text, one partitions the dynamics into intervals whose
duration is several times longer than the system's correlation time
(which is unity in our rescaled units). Then the intervals are
approximately statistically independent, and the probability of
reaching the target $X$ in each of the intervals is
proportional to $P_{\text{st}}\left(X\right)$, which is assumed
to be very small. Thus the FPT to reach the target $X$ obeys an
exponential distribution whose mean is given by
Eq.~\eqref{MFPTandSSD}, as stated in the main text. Relations such as
\eqref{MFPTandSSD} between MFPTs and SSDs (or quasi-SSDs) occur in
many situations when considering large deviations in stochastic
dynamical systems, see for instance Refs. \cite{WZKLT19, WKL20} in the
context of active matter, and \cite{MeersonAssaf2017} in the context
of stochastic population dynamics.  An even simpler example comes from
the Arrhenius law \eqref{Arrhenius} and Boltzmann distribution
\eqref{BoltzmannDist}.

\section{Donsker-Varadhan formalism}
\label{sec:DV}

Here, for completeness, we briefly outline the Donsker-Varadhan (DV) formalism \cite{DonskerVaradhan}, and apply it to the telegraphic noise in order to calculate the rate function $\Phi(z)$ that is given in Eq.~\eqref{Phidef} of the main text. For an accessible introduction and further details regarding the general DV theory, we refer the reader to Ref.~\cite{Touchette2018}.
Let $y\left(t\right)$ be a stochastic process with a finite state space, $y\left(t\right)\in\left\{ y_{1},\dots,y_{n}\right\} $, and let
\be
\frac{d\vect{v}}{dt}=L\vect{v}
\ee
be the master equation that describes the temporal evolution of the probability vector $\vect{v}=\left(v_{1},\dots,v_{n}\right)$, where $v_{i}(t)$ denotes the probability that $y\left(t\right)=y_i$. Here $L$, the generator of the dynamics, is an $n\times n$ matrix.
Define a dynamical observable
\be
A=\frac{1}{T}\int_{0}^{T}y\left(t\right)dt \, ,
\ee
and consider its PDF $p_{T}\left(a\right)$, defined via
\be
\text{Prob}\left(A\in\left[a,a+da\right]\right)=p_{T}\left(a\right)da
\ee
for infinitesimal $da$.
Then the DV formalism predicts that, in the limit $T \to \infty$, an LDP holds:
\be
p_{T}\left(a\right)\sim e^{-TI\left(a\right)} \, .
\ee
Moreover, it provides a method of calculating the rate function $I(a)$, by making use of the G\"{a}rtner-Ellis theorem \cite{Ellis}. From this theorem it follows that $I(a)$ is the Legendre-Fenchel transform
\be
I\left(a\right)=\max_{k\in\mathbb{R}}\left[ka-\mu\left(k\right)\right]
\ee
of the scaled cumulant generating function (SCGF)
\be
\mu\left(k\right)=\lim_{T\to\infty}\frac{1}{T}\left\langle e^{TkA}\right\rangle  \, .
\ee
Finally, the SCGF is given by the maximum eigenvalue of an auxiliary matrix operator
\be
L_{i,j}^{k}=\left(L^{\dagger}\right)_{i,j}+ky_{i}\delta_{i,j}\,,
\ee
which is a ``tilted'' version of the Hermitian conjugate of $L$. Here $\delta_{i,j}$ is the Kronecker delta.

Let us apply this formalism to the (rescaled) telegraphic noise $\sigma(t)$ that is defined in the main text. Its two possible states are $\sigma\left(t\right)\in\left\{ 1,-1\right\}$, and the generator of the dynamics of $\sigma(t)$ is the matrix
\be
L=\frac{1}{\tau_{c}}\left(\begin{array}{cc}
-1 & 1\\
1 & -1
\end{array}\right) \, .
\ee
Therefore, the SCGF is given by the largest eigenvalue of the matrix
\be
L^{k}=\frac{1}{\tau_{c}}\left(\begin{array}{cc}
-1 & 1\\
1 & -1
\end{array}\right)+k\left(\begin{array}{cc}
1 & 0\\
0 & -1
\end{array}\right) \, .
\ee
The largest eigenvalue of $L^k$ is
\be
\mu\left(k\right)=\sqrt{k^{2}+\frac{1}{\tau_{c}^{2}}}\,-\frac{1}{\tau_{c}} \, .
\ee
Since this function is convex, the Legendre-Fenchel transform becomes a Legendre transform, and we obtain
\be
I\left(a\right)=\frac{1-\sqrt{1-a^{2}}}{\tau_{c}} \, ,
\ee
recovering Eq.~\eqref{Phidef} of the main text.

%\medskip

\section{Langevin Dynamics simulations}
\label{app:simulations}

In order to verify the accuracy of Eqs.~\eqref{soflambda} and
\eqref{Xoflambda}, which constitute the analytical expression for the
LDF $s(X)$, we need to compute $P_{\rm st}(X)$, and then to extract
$s(X)$ from Eq.~(\ref{PstScaling}) [or practically, from the
  logarithmic version of it, Eq.~(\ref{LDPeq})]. The SSD can be
obtained by simulating an ensemble of long trajectories of particles
that follow the Langevin equation of motion
(\ref{LangevinDimensionless}), where for each trajectory an
independent realization of the telegraphic noise $\sigma(t)$ with a
characteristic switching time $\tau_c$ is randomly chosen. The
duration of the simulated trajectories $t$ must be sufficiently large
to ensure that steady state is established, while $\tau_c$ must be
sufficiently small in order to approach the limit $\tau_c\rightarrow
0$ of interest. Strictly speaking, in this straightforward approach,
the computed $P_{\rm st}(X)$ is expected to converge to the correct
SSD when the number of simulated trajectories becomes infinitely
large. In practice, however, this strategy fails in the large
deviation regime $X\lesssim X_0$ (which is of most interest) because
of the remarkably low probability of a trajectory to actually reach
near the edges of the support. To better appreciate the difficulty to
sample the large deviation regime, consider the case $\gamma=3$ [see
  Fig.~\ref{fig:sofXgamma3and1} (a)]. The first problem that we
encounter is the lack of a priori knowledge which $\tau_c$ is
``sufficiently close to zero''. A reasonable choice may be
$\tau_c=0.2$, which is smaller than the other (rescaled) time
constants relevant to the dynamics, i.e., (i) the inverse oscillator
frequency $\sqrt{k/m}=1$ and (ii) the friction damping time
$m/\Gamma=1/\gamma=1/3$. However, for this value of $\tau_c$, the
probability to be in the range $|X|>0.95$ [$s(X=0.95)\simeq 6.5$] is
smaller than $10^{-15}$, and the probability to reach $|X|>0.98$
[$s(X=0.98)\simeq 9$] is smaller than $10^{-21}$. These numbers
suggest that proper sampling of the large deviation regime may require
simulations of a prohibitively large number of
trajectories. Furthermore, even if we had the resources to do so, we
would still not know the difference between the LDF computed for a
small finite $\tau_c$ and the one expected in the limit
$\tau_c\rightarrow 0$.

Generally speaking, the number of times that the direction of the
noise changes in large deviations trajectories is considerably smaller
than the most probable number of switches, $n^*=t/\tau_c$, which is
characteristic of typical trajectories. In order to properly sample
such rare trajectories, we use the {\em exact}\/ relationship
\be
P(X)=\sum_{n=0}^{\infty}P(X|n)\Pi_n, \ee
where $P(X|n)$ is the conditional probability that the trajectory
ended at $x=X$ given that noise underwent $n$ switches during the
course of the motion, and $\Pi_n$ is the probability that the noise
changed sign exactly $n$ times. The latter is given by the Poisson
distribution \be
\Pi_n=\frac{e^{-t/\tau_c}}{n!}\left(\frac{t}{\tau_c}\right)^n.  \label{pnsum}
\ee
A trajectory with $n$ switching points consists of $n+1$ segments
during which the particle experiences a constant noise of magnitude
$\pm 1$. In order to generate such a trajectory, we draw $n+1$ random
numbers $R_i$ from a standard exponential distribution, and then
rescale them $R_i\rightarrow R_i(t/\sum_{j=1}^{n+1}R_j)$ to obtain the
durations of the dynamical segments. For a harmonic confining
potential, the equation of motion along each segment can be easily
solved, which yields the position and velocity of the particle at each
switching point. Once the trajectory has reached completion, the
position of the particle is recorded, and by simulating many
trajectories with $n$ switches, we obtain $P(X|n)$. An important merit
of this approach (besides the simulations of many rare trajectories)
is the fact that with the data for $P(X|n)$ (which depends on $t$ but
not on $\tau_c$), we can use Eq.~(\ref{pnsum}) to calculate the SSD
for any $\tau_c$ provided that the largest simulated $n$ is somewhat
larger than $n^*=t/\tau_c$. We can, thus, compute $s(X)$ for several
values of $\tau_c$, and then extrapolate the results to the limit
$\tau_c\rightarrow 0$.

\begin{figure*}[ht]
\includegraphics[width=0.46\linewidth,clip=]{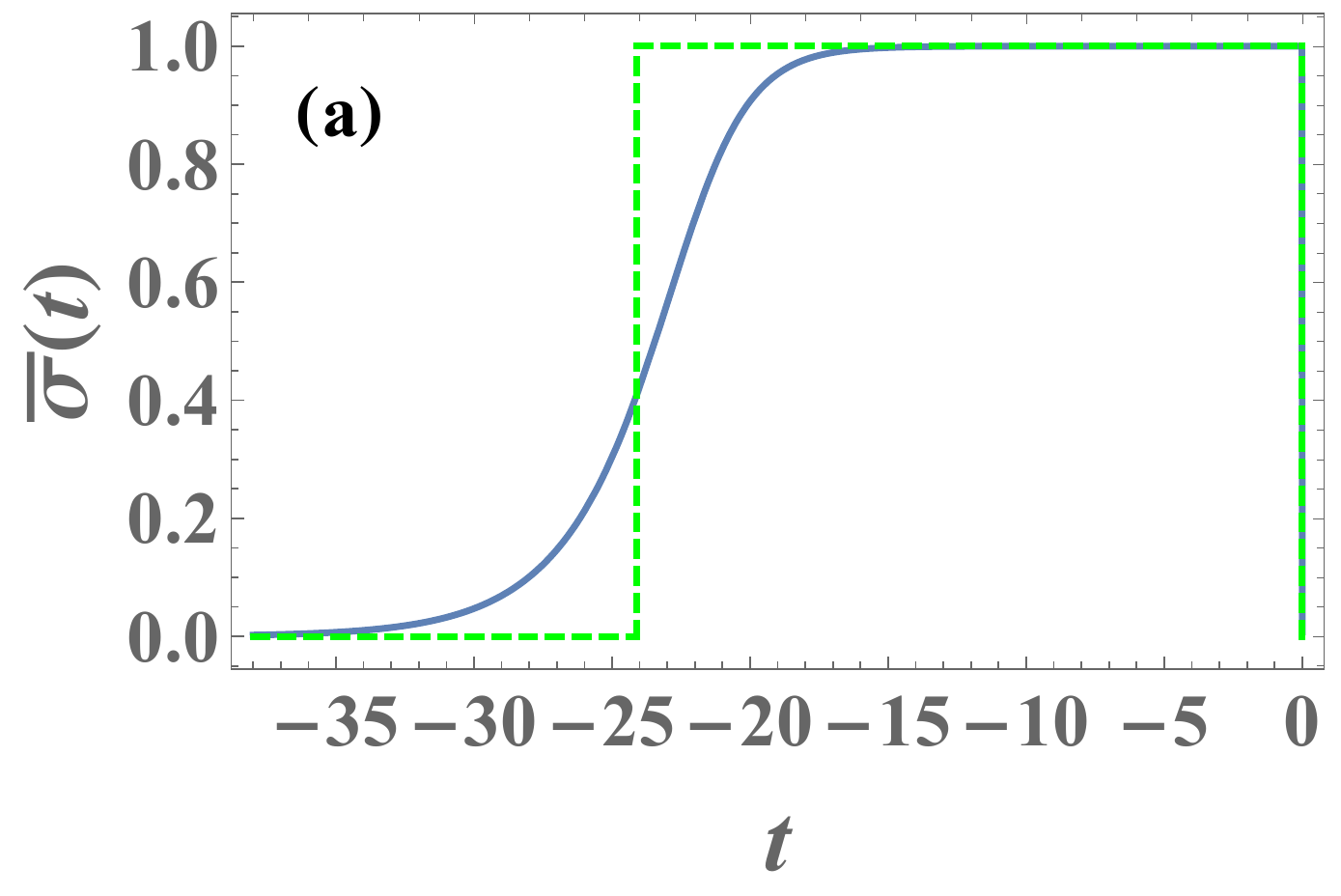}
\hspace{1mm}
\includegraphics[width=0.48\linewidth,clip=]{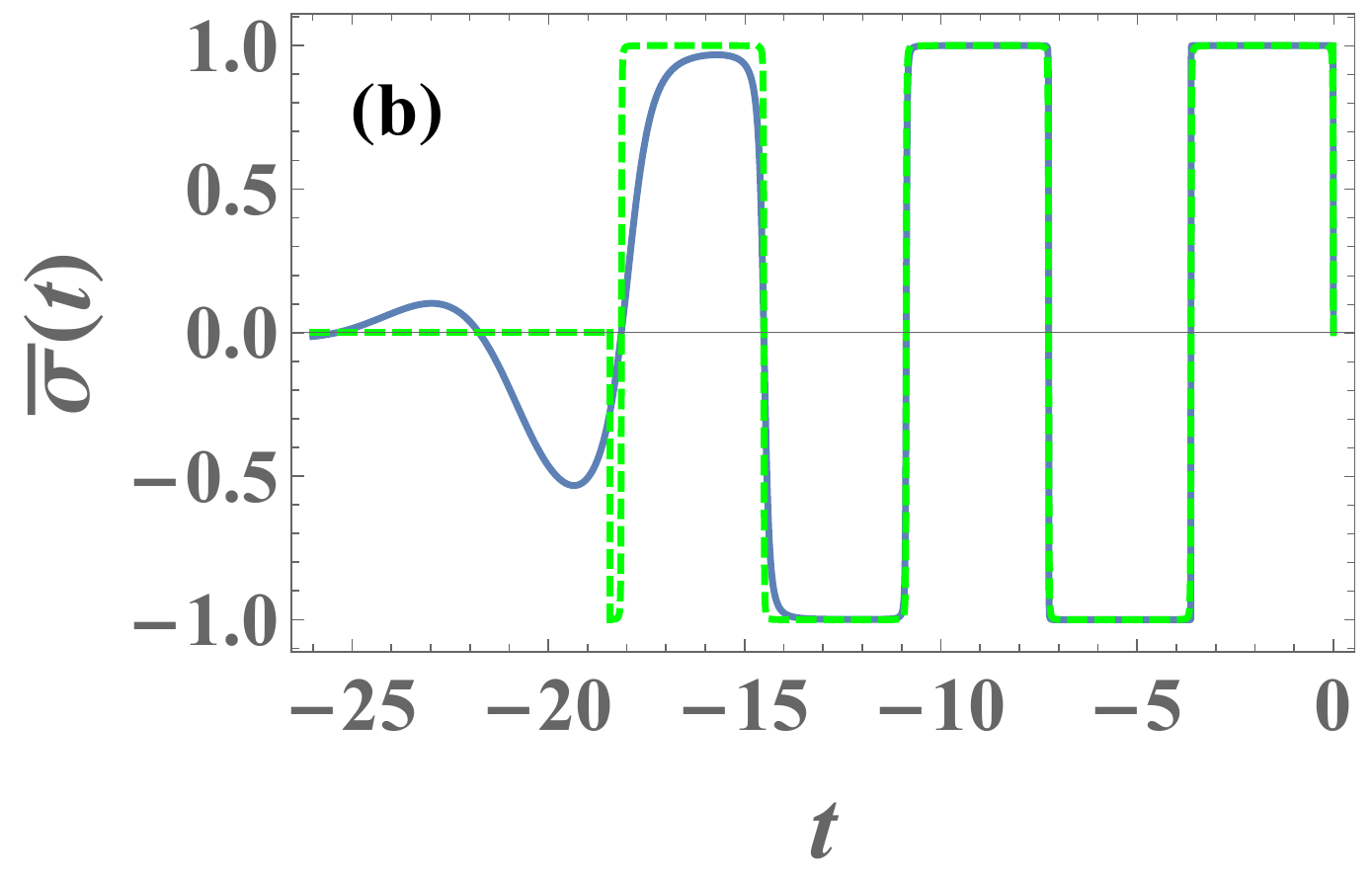}
\caption{The optimal realization $\bar{\sigma}\left(t\right)$ of the
  coarse-grained noise term, $\bar{\sigma}(t)$, in the limit $\lambda
  \gg 1$, in the overdamped (a) and underdamped (b) cases.  Solid
  lines correspond to the exact result \eqref{Psol}.  Dashed lines
  correspond to the approximate expressions, \eqref{Papprox} and
  \eqref{PapproxUD} respectively.  Parameters are $\gamma=3$ and
  $\lambda=10^{4}$, corresponding to $X = 0.999738\dots$ in (a), and
  $\gamma=1$ and $\lambda=10^{4}$, corresponding to $X =1.38938\dots$
  in (b). Recall that in (a) the edge of the support of the SSD is at
  $X_0 = 1$ while in (b), it is at $X_0 =
  \coth\left(\frac{\pi}{2\sqrt{3}}\right)=1.38958\dots$.}
\label{fig:Poft}
\end{figure*}

The results presented in Fig.~\ref{fig:sofXgamma3and1} are based on
simulations of $10^9$ trajectories of duration $t=10$, for values of
$n$ in the range $0\leq n\leq275$. The displayed $s(X)$ is an
extrapolation of the LDFs calculated for $\tau_c=0.2$, 0.3, 0.4, and
0.5. 
%For $\gamma=3$, the agreement with the analytical prediction is better than 5\% for values of $|X|$ as large as 0.96. 
For $\gamma=3$, the computational data is indistinguishable from the analytical predictions at $|X|<0.92$, and at $0.92<|X|<0.96$ the difference is at most 5\%.
We note that:
(i) Simulating straightforwardly (with an overall similar CPU time)
the dynamics with $\tau_c=0.2$ yielded no single trajectory with
$|X|>0.87$. (ii) In a future publication we will demonstrate how the
approach presented herein can be further improved.

%\medskip

\section{Solving the integral for $X_0$ in the underdamped regime}
\label{app:X0}

In this appendix we solve the integral that appears in the first line
of Eq.~\eqref{X0sol} of the main text, thereby obtaining the second
line in that equation.  Denoting $\omega \equiv \sqrt{4-\gamma^{2}} \,
/ 2$, the first line of Eq.~\eqref{X0sol} can be rewritten as
$X_{0}=I/\omega$ where \be
\label{I1}
I=\int_{-\infty}^{0}e^{\gamma t/2}\left|\sin\left(\omega t\right)\right|dt \, .
\ee
We now divide the integration region into two parts, and obtain
\bea
\label{I2}
&&I = \int_{-\infty}^{-\pi/\omega}e^{\gamma t/2}\left|\sin\left(\omega t\right)\right|dt+\int_{-\pi/\omega}^{0}e^{\gamma t/2}\left|\sin\left(\omega t\right)\right|dt \nn\\
&&=e^{-\pi\gamma/2\omega}\int_{-\infty}^{0}e^{\gamma t/2}\left|\sin\left(\omega t\right)\right|dt-\int_{-\pi/\omega}^{0}e^{\gamma t/2}\sin\left(\omega t\right)dt , \nn\\
\eea
where, when moving from the first to the second line in
Eq.~\eqref{I2}, we shifted the integration variable $t\to
t+\pi/\omega$ in the first integral, and used $\sin\left(\omega
t\right) < 0$ in the second integral.  Noticing now that the first
integral in the second line of Eq.~\eqref{I2} coincides exactly with
the definition \eqref{I1} of $I$, we then solve Eq.~\eqref{I2} for $I$
to get \bea
\label{I3}
I&=&-\frac{1}{1-e^{-\pi\gamma/2\omega}}\int_{-\pi/\omega}^{0}e^{\gamma t/2}\sin\left(\omega t\right)dt\nn\\
&=&\frac{4\omega}{\left(\gamma^{2}+4\omega^{2}\right)}\coth\left(\frac{\pi\gamma}{4\omega}\right)=\omega\coth\left(\frac{\pi\gamma}{4\omega}\right) \, ,
\eea
where, in the last equality, we used the definition of $\omega$.
Finally, using $X_0 = I/\omega$ with Eq.~\eqref{I3} and the definition
of $\omega$, we arrive at the second line of Eq.~\eqref{X0sol} that is
given in the main text.

%\medskip

\section{Near the edges of the support}
\label{app:nearEdges}

In order to check our approximations for $\bar{\sigma}$, given in
Eqs.~\eqref{Papprox} and \eqref{PapproxUD} of the main text, we
plotted them in Figure \ref{fig:Poft}, together with the corresponding
exact solutions for $\bar{\sigma}\left(t\right)$ from Eq.~\eqref{Psol}
for a large value of $\lambda$, in the overdamped (a) and underdamped
(b) cases, respectively.  The agreement is very good.  There are
temporal boundary layers, at $t\simeq t_{\lambda}$ and at $t\simeq 0$,
and also around times at which the sign of $\bar{\sigma}$ changes in
the underdamped case. However, their relative width of these boundary
layers vanishes in the large-$\lambda$ limit.  The approximation
improves as $\lambda$ is increased (not shown).

%\medskip

\section{The strongly-overdamped limit}
\label{Appendix:OD}

The exact solution of the integral in Eq.~\eqref{soflambdaOD1} yields
%\be
%s \simeq \frac{\gamma}{2}\left[\ln\left(\sqrt{1+\frac{\lambda^{2}}{\gamma^{2}}}+1\right)-\ln\left(\sqrt{1+\frac{\lambda^{2}}{\gamma^{2}}}-1\right)-2\ln2+2\ln\left|\frac{\lambda}{\gamma}\right|\right]
%\ee
\bea
\label{soflambdaOD2}
s &\simeq& \frac{\gamma}{2}\left[\ln\left(\sqrt{1+\frac{\lambda^{2}}{\gamma^{2}}}+1\right)-\ln\left(\sqrt{1+\frac{\lambda^{2}}{\gamma^{2}}}-1\right)\right. \nn\\
&&\left.-2\ln2+2\ln\left|\frac{\lambda}{\gamma}\right|\right] \, .
\eea
Now, using Eq.~\eqref{dsdlambda} with \eqref{soflambdaOD2}, we find
\be
\frac{dX}{d\lambda}=\frac{1}{\lambda}\frac{ds}{d\lambda} \simeq \frac{\gamma}{\lambda^{2}}\left(1-\frac{\gamma}{\sqrt{\gamma^{2}+\lambda^{2}}}\right)
\ee
which we immediately integrate, together with the boundary condition $X(\lambda=0) = 0$, to obtain
\be
X \simeq \frac{\sqrt{\gamma^{2}+\lambda^{2}}-\gamma}{\lambda} \, .
\ee
Finally, we invert this relation,
\be
\lambda \simeq \frac{2\gamma X}{1-X^{2}} \, ,
\ee
which, after plugging into Eq.~\eqref{soflambdaOD2}, yields 
\be
s(X) \simeq - \gamma\ln\left(1-X^{2}\right) \, ,
\ee
coinciding with Eq.~\eqref{sOverDamped} of the main text.


\begin{thebibliography} {99}
\bibitem{footnote:thermalNoise}
Note, however, that it is easy to take into account an additional thermal noise, as we explain in the Discussion (section \ref{sec:conclusion}), see also Ref.~\cite{SLMS22}.
\bibitem{footnote:whiteNoise} The white-noise limit is recovered in the joint limit $\mathcal{T}_{c}\to 0$, $\Sigma_{0} \to \infty$, such that $\Sigma_{0}^{2}\mathcal{T}_{c}$ is held constant.
\end{thebibliography}
\end{document}